\documentstyle[epsfig,graphics]{mn}

\begin{document}

\newcommand{\Zsolar}{\mbox{\,$\rm Z_{\odot}$}}
\newcommand{\etal}{{et al.}\ }
\newcommand{\ang}{\mbox{$\rm \AA$}}
\newcommand{\xs}{$\chi^{2}$}
\newcommand{\be}{\begin{equation}}   
\newcommand{\ee}{\end{equation}}     

\title[The SCUBA 8-mJy survey]{The SCUBA 8-mJy survey - I:
Sub-millimetre maps, sources and number counts.}

\author[S.E. Scott et al..]
{S.E. Scott$^{1}$, M.J. Fox$^{2}$, J.S. Dunlop$^{1}$, S. Serjeant$^{2,3}$,
J.A. Peacock$^{1}$, R.J. Ivison$^{4}$, \and
S. Oliver$^{5}$, R.G. Mann$^{1}$,
A. Lawrence$^{1}$, A. Efstathiou$^{2}$, M. Rowan-Robinson$^{2}$, \and
D.H. Hughes$^{6}$, E.N. Archibald$^{7}$, A. Blain$^{8}$, M. Longair$^{9}$  
\\
$^{1}$Institute for Astronomy, University of Edinburgh, Royal Observatory,
Blackford Hill, Edinburgh, EH9 3HJ, UK\\
$^{2}$Astrophysics Group, Blackett Laboratory, Imperial College, 
Prince Consort Rd., London SW7 2BW, UK\\
$^{3}$Unit for Space Sciences and Astrophysics, School of Physical
Sciences, University of Kent, Canterbury, Kent CT2 7NZ, UK\\
$^{4}$UK ATC, Royal Observatory,Blackford Hill, Edinburgh, EH9 3HJ, UK\\
$^{5}$Astronomy Centre, CPES, University of Sussex, Falmer, Brighton
BN1 9QJ, UK\\
$^{6}$Instituto Nacional de Astrof\'{\i}sica, \'{O}ptica y Electr\'{o}nica
(INAOE), Apartado Postal 51 y 216, 72000 Puebla, Pue., Mexico\\
$^{7}$Joint Astronomy Centre, 660 N. A'ohoku Place, Hilo, Hawaii
96720, USA\\
$^{8}$Institute of Astronomy, University of Cambridge, Madingly Road,
Cambridge CB3 0HE, UK\\
$^{9}$Cavendish Astrophysics Group, Cavendish Laboratory, Madingley
Road, Cambridge CB3 0HE, UK}

\date{Published in MNRAS, 331, 817}

\maketitle
  
\begin{abstract}
We present maps, source lists, and derived number counts from the largest, 
unbiassed, extragalactic sub-mm survey so far undertaken with the SCUBA camera 
on the JCMT. Our maps are located in two regions of sky (ELAIS N2 and 
Lockman-Hole E) and cover 260 arcmin$^2$, to a typical rms noise level of 
$\sigma_{850} \simeq 2.5$~mJy/beam. We have reduced the data using both the 
standard JCMT SURF procedures, and our own IDL-based pipeline which produces 
zero-footprint maps and noise images. The uncorrelated noise maps produced 
by the latter approach have enabled us to apply a maximum-likelihood method 
to measure the statistical significance of each peak in our maps, leading to 
properly-quantified errors on the flux density of all potential sources.

We detect 19 sources with S/N~$> 4$, and 38 with S/N~$> 3.5$. To assess both the completeness of this survey, and the impact of 
source confusion as a function of flux density we have applied our 
source-extraction algorithm to a series of simulated images. The result is a 
new estimate of the sub-mm source counts over the flux-density range $S_{850} 
\simeq 5 - 15$~mJy, which we compare with estimates derived by other workers, 
and with the predictions of a number of models. 

Our best estimate of the cumulative source count at $S_{850} > 8$~mJy is 
$320^{+80}_{-100}$ per square degree. Assuming that the majority of sources 
lie at $z > 1.5$, this result implies that the co-moving number density of 
high-redshift galaxies forming stars at a rate in excess of 1000$~{\rm 
M_{\odot}yr^{-1}}$ is $\simeq 10^{-5}~{\rm Mpc^{-3}}$, with only a weak 
dependence on the precise redshift distribution. This number density 
corresponds to the number density of massive ellipticals with $L > 3-4 
L^{\star}$ in the present-day universe, and is also the same as the co-moving 
number density of comparably massive, passively-evolving objects in the 
redshift band $1 < z < 2$ inferred from recent surveys of extremely red 
objects. Thus the bright sub-mm sources uncovered by this survey can 
plausibly account for the formation of all present-day massive ellipticals. 
Improved redshift constraints, and ultimately an improved measure of sub-mm 
source clustering can refine or refute this picture. 
\end{abstract}


\newpage

\vspace*{1cm}

\section{Introduction}

Over the past three years deep sub-millimetre surveys, carried out to
a variety of depths (Smail et al. 1997, Hughes et al. 1998,
Barger, Cowie \& Sanders 1999, Blain et al. 1999, Eales et al. 2000, Borys et al. 2002) have highlighted 
the extreme importance of dust in the
determination of the global star-formation history of the
Universe. Optical/UV studies originally suggested a steep rise in
the star-formation rate (SFR) as a function of redshift between $z=0$
and $z=1$ (Lilly et al. 1996), peaking at $1.0<z<1.5$, and declining to
values comparable to those observed in the present day at $z \simeq 4$
(Madau et al. 1996). More recent optical/UV analyses have implied a
much gentler trend in SFR density with epoch, Cowie et al. (1999) finding a
more gradual slope ($\sim (1+z)^{1.5}$
for an Einstein de Sitter cosmology) at low redshift
($z<1$) than had previously been determined
by Lilly et al. (1996), and at high redshift
the Lyman break population (Steidel et al. 1999) suggesting an
approximately uniform value for $1<z<4$. However, in heavily dust-enshrouded star-forming
regions, much of the optical/UV radiation emitted by the 
young stars is absorbed, leading to the possibility that a
significant amount of star formation, particularly at high redshift, 
may have been missed in these wavebands. 

In order to quantify the star-formation density contributed by such 
highly-obscured objects, we must observe directly the 
rest-frame far-infrared (FIR) emission from the heated dust, which at
redshifts greater than $z \sim 1$ is shifted into the
sub-millimetre waveband. The steep spectral index of the thermal emission
longwards of the peak at $\lambda \simeq 100$~${\rm \mu m}$ 
results in a sufficiently large negative
K-correction that the dimming effect of increasing cosmological
distance is effectively cancelled over a wide range in redshift. 
Consequently, for a galaxy with fixed intrinsic FIR luminosity, we
would expect to observe approximately the same 850~${\rm \mu m}$ flux
density for an object at $z=8$ as at $z=1$. The strength of this
effect is obviously dependent on the assumed cosmology. It is most
pronounced in an Einstein-de Sitter universe, in which the predicted 
flux density of an object of fixed luminosity actually increases
slightly beyond $z \simeq 1$. 
If instead we adopt $\Omega_M = 0.3$, $\Omega_{\Lambda} = 0.7$, 
a very gentle decline in flux density is predicted over 
this same redshift range.

Indications that a large fraction of star-forming activity at
high redshift is heavily enshrouded in dust have been provided, for example,
by a deep sub-millimetre survey, of the northern HDF (Hughes et al.
1998). The discovery of 5 discrete sources with either 
very faint or presently undetected optical identifications, accounting for
approximately $30-50\%$ of the sub-millimetre background observed by
COBE-FIRAS (Puget et al. 1996, Fixsen et al. 1998, Hauser et al. 1998), have suggested
a star-formation density at least a factor of $\simeq 5$ higher at $z>2$
than that originally deduced from optical/UV data. The results of
Barger, Cowie \& Richards (2000) and Eales et al. (2000) are also
consistent with a high redshift SFR density up to an order of magnitude higher
than that inferred in the optical/UV. These sources, are believed to be
analogous to the local ultraluminous infrared galaxies (ULIRGs) and
undergoing a period of massive star-formation at rates of $10^{2} -
10^{3} \mathrm{M_{\odot}yr^{-1}}$.

The discovery of such objects at high redshift has important
implications for the formation mechanism of massive elliptical
galaxies. In current cold dark matter (CDM) based models, elliptical
galaxies are created at modest redshift by the merging of
intermediate-mass discs (Baugh, Cole \& Frenk 1996). Although some
recent semi-analytic results appear supportive of the formation of
massive ellipticals at modest redshift by means of hierarchical mergers
(Kauffman \& Charlot 1998), observations continue to imply that a massive
coeval starburst of timescale $\simeq 1$ Gyr at high redshift ($z>3$)
is required to explain the properties of at least some ellipticals
(Jimenez et al. 1999, Renzini \& Cimatti 2000). Additionally, the
connection between black-hole and spheroid mass (Magorrian et al. 1998,
Kormendy \& Gebhardt 2001)
suggests that galaxies hosting an active nucleus may be more
representative of the general population than had previously been
thought, and might indicate a link between spheroid formation and the
epoch of maximum quasar activity at $z=2-3$.

The discovery of the Lyman-break population at $z=2-4$ (Steidel et al. 1999)
has prompted the suggestion that the formation of present-day galactic
bulges and ellipticals has already been observed at optical
wavelengths. However, the most massive ellipticals contain $10^{12}$
solar masses of stars, requiring a sustained SFR of 
$\sim 1000~{\rm M_{\odot}yr^{-1}}$ for formation 
over a 1 Gyr period. Even if the
most optimistic corrections are made for the effects of dust, the
inferred star formation rates in these objects (typically 
$3-30~\mathrm{M_{\odot}yr^{-1}}$) fall short by $1-2$ orders of
magnitude. Moreover, the first direct tests of such corrections
suggest they may indeed have been exaggerated. Direct measurements of the
sub-millimetre/millimetre emission of the strongly-lensed Lyman-break
galaxies MS 1512+36-cB58 (van der Werf et al. 2001, Baker et al. 2001,
Sawicki 2000) and MS 1358+62-G1 (van der Werf et al. 2001)
imply that the procedure used to predict the sub-millimetre
emission of the Lyman-break population over-predicts the observed 
850~${\rm \mu m}$ flux densities by a factor of up to 14. 
A recent X-ray detection of MS 1512+36-cB58 (Almaini, Pettini \&
Steidel in prep) also supports the view that corrections for dust have
been exaggereated, the level of X-ray emission implying a much lower SFR than
that derived from the ``dust-corrected'' UV spectrum, but consistent
with the
measured 850~${\rm \mu m}$ flux density, for a typical starburst SED. 

One possibility is that spheroid formation is too widely distributed
in space and/or time to give rise to such spectacular starbursts.
However, observationally we may be missing massive starbursts in
the optical as a result of dust obscuration or because they are at too
great a redshift. Sub-millimetre detections of the $z \simeq
4$ radio galaxies 4C41.17 (Dunlop et al. 1994, Hughes, Dunlop \&
Rawlings 1997) and 8C1435+635 
(Ivison 1995, Ivison et al. 1998) confirm that the Lyman-break population 
does not tell the full story. These extreme radio galaxies display the 
properties expected of youthful massive ellipticals, and have inferred dust 
masses $> 10^{8} \mathrm{M_{\odot}}$ and SFRs 
$>1000~{\rm M_{\odot}yr^{-1}}$.  

Here we present the first results from the `SCUBA 8~mJy Survey', which is 
a wider-area, somewhat shallower survey than its
earlier counterparts (HDF-N - Hughes et al. 1998; Clusters Survey -
Smail et al. 1997, Blain et al. 1998; Hawaii Fields - Barger, Cowie \&
Sanders 1999; CUDSS - Eales et al. 2000), undertaken with the aim of constraining the
brighter end of the 850~${\rm \mu m}$ source counts in the region of 8~mJy. 
This survey has three key
advantages over previous sub-millimetre surveys. Firstly, the 
relatively bright flux-density limit
facilitates follow-up with existing instrumentation which should
eventually yield an accurate astrometric position for every source via
either the VLA, or in the case of the very high redshift sources
($z>3-4$) IRAM PdB. Secondly, the steep slope of the source counts
makes confusion a serious issue when probing deeper than $\sim 2-3$~mJy. The number density of sources brighter than 8~mJy is a factor of
$\simeq 3$ lower than at 3~mJy and so source-confusion is much less of
a problem in this survey. Finally, any sources discovered in this
survey must be as bright or brighter in the sub-millimetre than the
extreme radio galaxies mentioned above. This survey is therefore
optimally placed for detecting the most luminous starbursts with SFR
$\simeq 1000~{\rm M_{\odot}yr^{-1}}$, and through an extensive
follow up program already underway (Fox et al. 2002, Lutz et al.
2001) will provide clues to the formation of the most massive ellipticals.  

This paper is laid out as follows. In Sections 2 and 3 we describe the
observations and data reduction. In Section 4 we outline the source
extraction algorithm used to create our source catalogues. This is
described in full mathematical detail in Appendix A. In Section 5
we present and discuss the results of simulations carried out both in
conjunction with the real data, and through the production of
fully-simulated survey areas. In Section 6 we present our source
lists, giving positions, along with 850 and 450~${\rm \mu m}$ flux densities. 
In Section 7 we derive the resulting
source counts (both raw and corrected using the simulation results). In
section 8 we present 2-point auto-correlation functions for each of
our survey fields. A
full discussion of the results, including comparison with other
surveys, calculated dust masses, star-formation rates, estimates of co-moving 
number density, and clustering properties is given in Section
9. Finally, our conclusions are summarised in Section 10. In subsequent
papers, the cross-correlation of our SCUBA sources with deep X-ray and
radio observations will be discussed by Almaini et al. (2002) and Ivison
et al. (2002) respectively.

For the calculation of physical quantities we have assumed 
$H_0 = 67$~${\rm km s^{-1}Mpc^{-1}}$ 
throughout, and all cosmological calculations
have been performed for both an Einstein-de Sitter universe and 
a $\Lambda$-dominated flat cosmology with 
$\Omega_M = 0.3$, $\Omega_{\Lambda} = 0.7$.

\begin{table*}
\begin{tabular}{|c|r|c|c|c|c|} \hline
 & & ELAIS N2  & ELAIS N2 & Lockman Hole East & Lockman Hole East \\
 & & Uniform noise & Full area & Uniform noise & Full area \\ \hline
Area (sq. arcmin) &      & 63 & 102 & 55 & 122 \\ \hline
\% Differential   & 5~mJy & $12.3 \pm 4.6\phantom{0}$ & \phantom{0}$8.4 \pm 3.2\phantom{0}$ & $37.1 \pm 10.3$ & $20.5 \pm 5.1$ \\
completeness & 8~mJy & $53.8 \pm 10.2$ & $34.1 \pm 6.1\phantom{0}$ & $68.3 \pm 12.9$ & $45.5 \pm 7.7\phantom{0}$ \\
at flux density & 11~mJy & $84.5 \pm 12.1$ & $62.4 \pm 8.2\phantom{0}$ & $82.9 \pm 15.4$ & $47.8 \pm 7.3\phantom{0}$ \\
limit & 14~mJy & $98.3 \pm 12.8$ & $74.0 \pm 8.8\phantom{0}$ & $97.7 \pm 15.1$ & $73.0 \pm 9.1\phantom{0}$ \\ \hline
\% Integral & 5~mJy & $45.4 \pm 2.1\phantom{0}$ & $30.8 \pm 1.3\phantom{0}$ & $50.2 \pm 2.5\phantom{0}$ & $31.6 \pm 1.4\phantom{0}$ \\
completeness & 8~mJy & $79.5 \pm 3.2\phantom{0}$ & $57.8 \pm 2.1\phantom{0}$  & 
$80.7 \pm 3.8\phantom{0}$ & $52.2 \pm 2.0\phantom{0}$ \\
to flux density & 11~mJy & $92.0 \pm 4.3\phantom{0}$ & $68.8 \pm 2.9\phantom{0}$ & 
$92.3 \pm 5.2\phantom{0}$ & $62.2 \pm 2.8\phantom{0}$ \\
limit & 14~mJy & $98.2 \pm 7.7\phantom{0}$ & $73.0 \pm 5.1\phantom{0}$ & 
$96.0 \pm 8.8\phantom{0}$ & $72.4 \pm 5.2\phantom{0}$  \\ \hline
Boosting & 5~mJy & $1.28 \pm 0.05$ & $1.28 \pm 0.05$ & $1.35 \pm 0.06$ & $1.59 \pm 0.14$ \\
factor & 8~mJy & $1.16 \pm 0.04$ & $1.20 \pm 0.04$ & $1.12 \pm 0.04$ & $1.15 \pm 0.04$ \\
at flux density & 11~mJy & $1.06 \pm 0.03$ & $1.08 \pm 0.03$ & $1.10 \pm 0.03$ & $1.14 \pm 0.03$ \\
limit & 14~mJy & $1.02 \pm 0.02$ & $1.04 \pm 0.02$ & $1.03 \pm 0.03$ & $1.05 \pm 0.02$ \\ \hline
Positional & 5~mJy & $2.42 \pm 0.57$ & $2.42 \pm 0.57$ & $4.52 \pm 0.53$ & $4.85 \pm 0.49$ \\
uncertainty & 8~mJy & $3.63 \pm 0.34$ & $3.92 \pm 0.38$ & $3.87 \pm 0.47$ & $3.82 \pm 0.40$ \\
at flux density & 11~mJy & $2.53 \pm 0.27$ & $2.60 \pm 0.23$ & $2.81 \pm 0.28$ & $2.96 \pm 0.24$ \\
limit /arcsec& 14~mJy & $2.39 \pm 0.18$ & $2.65 \pm 0.18$ & $2.17 \pm 0.17$ & $2.45 \pm 0.18$  \\ \hline
\end{tabular}
\label{table:compsim1}\caption{\small Results of the completeness
 simulations at the $3.50\sigma$ significance level, in which
 individual sources of the specified flux density given in column 2 were added
 into the unconvolved survey data taken up to and including 16th
 August 2000, and retrieved by means of the source
 extraction algorithm outlined in Section 4.}
\end{table*}

\section{Observations}
The survey is divided between two fields; the Lockman Hole East
(centred at RA 10:52:8.82, DEC +57:21:33.8) and
ELAIS N2 (centred at RA 16:36:48.85, DEC +41:01:48.5). These areas both have a vast quantity of multi-wavelength
data available for follow-up studies, and were selected to
coincide with deep Infrared Space Observatory (ISO)
surveys at 6.7, 15, 90 and 175~${\rm \mu m}$ (Lockman Hole East -- Elbaz et
al. 1999, Kawara et al. 1998; ELAIS N2 -- Oliver et al. 2000, Serjeant
et al. 2000, Efstathiou et al. 2000).

Pilot observations of the Lockman Hole East and ELAIS N2 fields began in
March 1998 and July 1998 respectively, using the Sub-millimetre Common
User Bolometer Array (SCUBA) (Holland et al. 1999) on the James Clerk Maxwell Telescope
(JCMT) on Mauna Kea, Hawaii. In brief, SCUBA consists of two
feedhorn-coupled bolometer arrays,
both of which are arranged in a close-packed hexagon and have
approximately a 2.3 arcmin field of view (slightly smaller on the
short-wavelength array). The diffraction-limited beams delivered
by the JCMT have FWHM of 14.5$^{\prime \prime}$ at 
850~${\rm \mu m}$ and 7.5$^{\prime \prime}$ at 450~${\rm \mu m}$,
and the
bolometer feedhorns on the arrays are sized for optimal coupling to
the respective beams. As a result, the long-wavelength (LW) array
consists of 37 bolometers, and the short-wavelength (SW) array
consists of 91 bolometers. Observations may be made simultaneously
with both arrays by means of a dichroic beamsplitter. A
$^{3}$He/$^{4}$He dilution refrigerator cools SCUBA down to $\sim$~90~mK, 
making observations with the array sky-background limited.

The most efficient mapping technique for sources smaller than the
field of view is `jiggle-mapping', and is the method employed in this
survey. As explained above, the bolometers are arranged 
such that the sky is under-sampled
at any instant and so the secondary is `jiggled' in a hexagonal
pattern to fill in the gaps. In order to fully sample the sky at both
wavelengths a 64-point jiggle-map is required. Temporal variations and
linear spatial gradients in the sky were removed by chopping and
nodding. A 30-arcsec chop throw was chosen in order to optimise the
sky subtraction. The chop
direction is fixed in celestial co-ordinates, lying directly north-south in
the Lockman Hole East, and 48$^{\circ}$ east of north in ELAIS N2,
parallel to the long axis of the survey area (to ensure the chop
remains within the boundaries).

The preliminary observing strategy was to build up a linear strip in
each field by overlapping SCUBA jiggle-maps, staggered by half the width of the
array. The data collected from the pilot Lockman Hole East run
in March 1998 were composed of a series of jiggle-maps, forming a
deep strip at the centre of the survey area with $\sigma_{850} \simeq
1.6$~mJy/beam. In subsequent observing runs, the strategy was modified to that
of an overlapping `tripod' positioning scheme, with three offset grids
of hexagonally-close-packed pointings, and the overall
integration time reduced by a factor of $\sim 2$ to make more
efficient use of the allocated observing time. Each tripod position
received 3.0 hours of integration time, yielding
a typical noise level of $\sigma_{850} \simeq 2.5$~mJy/beam. In
order to give near homogeneous sky noise across the final coadded map,
high and low airmasses were evenly distributed throughout. We
have mapped a total area of 260~$\mathrm{arcmin^{2}}$,
up to and including observations made in March 2001. 

Prior to 15th December 1999, the observations were made using the
narrow-band 850/450~${\rm \mu m}$ filters. The majority of the
observations carried out after this date made use of the
more sensitive wide-band 850/450~${\rm \mu m}$ filters. Since
the central wavelength of the 850~${\rm \mu m}$ filter did not change by
more than 1~${\rm \mu m}$, the data taken with the wide- and narrow-band
filters were directly combined in producing the final images.
  
Skydips and pointings were carried out, on average, every $1.5-2$
hrs. The observations were carried out in dry weather conditions,
under the constraint $\tau_{225 \mathrm{GHz}} < 0.08$. 
Primary/secondary calibration observations were taken at the start
and end of every shift, with additional beam maps of the pointing source
(0923+392 in the case of the Lockman Hole East, and 3C345 in the case
of ELAIS N2) being made every 3-4 hours to provide tertiary 
flux-density calibration. 

\begin{figure*}
\begin{centering}
\centerline{\epsfig{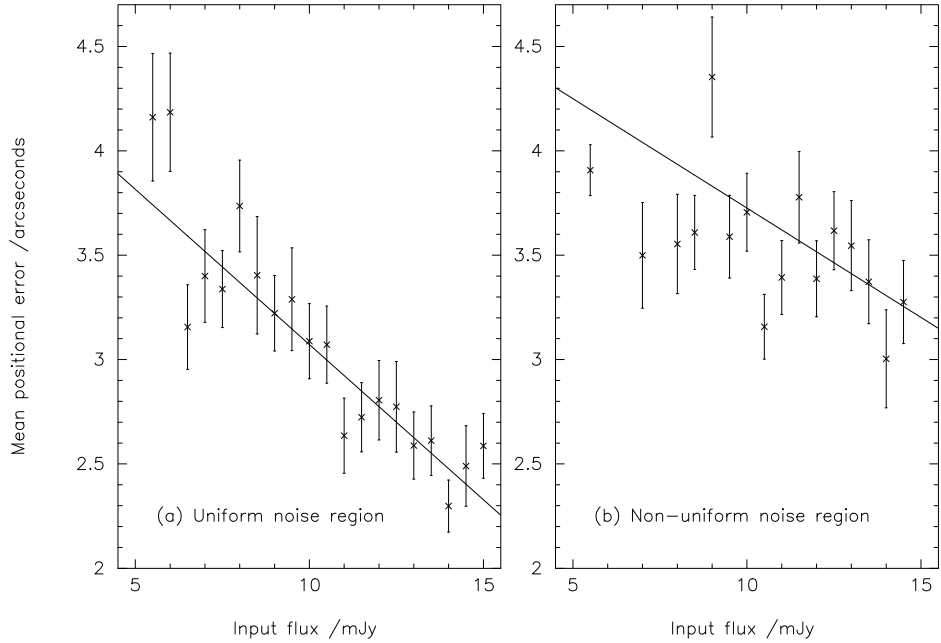}}
\caption{\small{The dependence of positional error on input source
flux density in the range $5-15$~mJy, based on a single template
source repeatedly being added into and then retrieved ($>3.50\sigma$) from the real unconvolved
signal maps. The error bars give the standard error on the
mean. Plot (a) gives the results from the uniform noise region,
whereas plot (b) gives the results for the non-uniform edge regions
which have not received the full integration time. The solid lines are a
best-fit to the plotted data points.}
\label{fig:pos_err}} 
\end{centering}

\end{figure*}
\begin{figure*}
\begin{centering}
\centerline{\epsfig{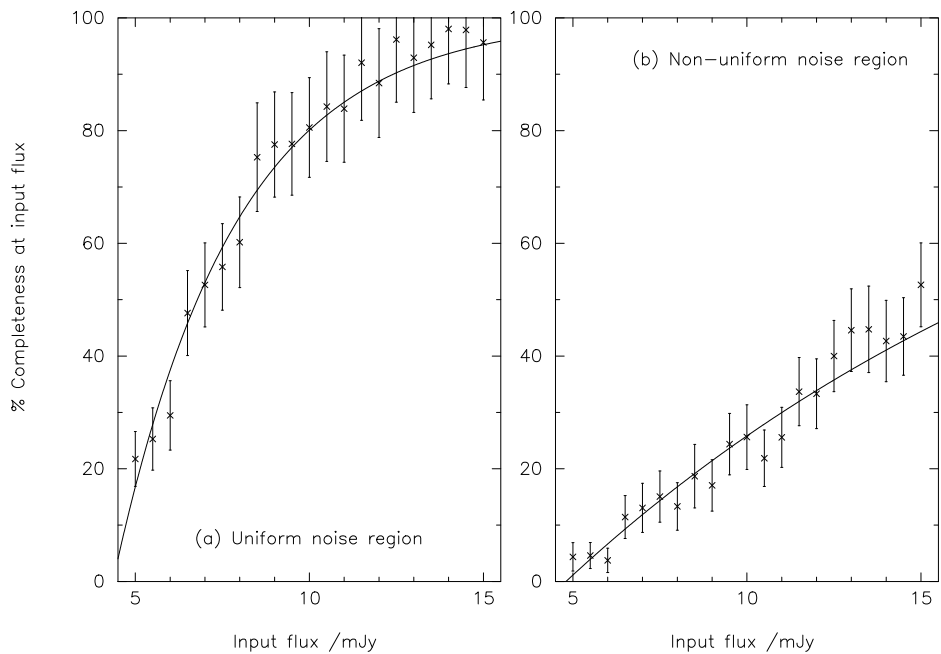}}
\caption{\small{The percentage of sources returned against input source
flux density in the range $5-15$~mJy, based on a single template
source repeatedly being added into 
and retrieved ($>3.50\sigma$) from the real unconvolved
signal maps. The error bars give the standard error on the
mean. Plot (a) gives the results from the uniform noise region,
whereas plot (b) gives the results for the non-uniform edge regions
which have not received the full integration time. The solid curves show
a best-fit exponential-based model to the plotted data points. }
\label{fig:comp}} 
\end{centering}
\end{figure*}

\begin{figure*}
\begin{centering}
\centerline{\epsfig{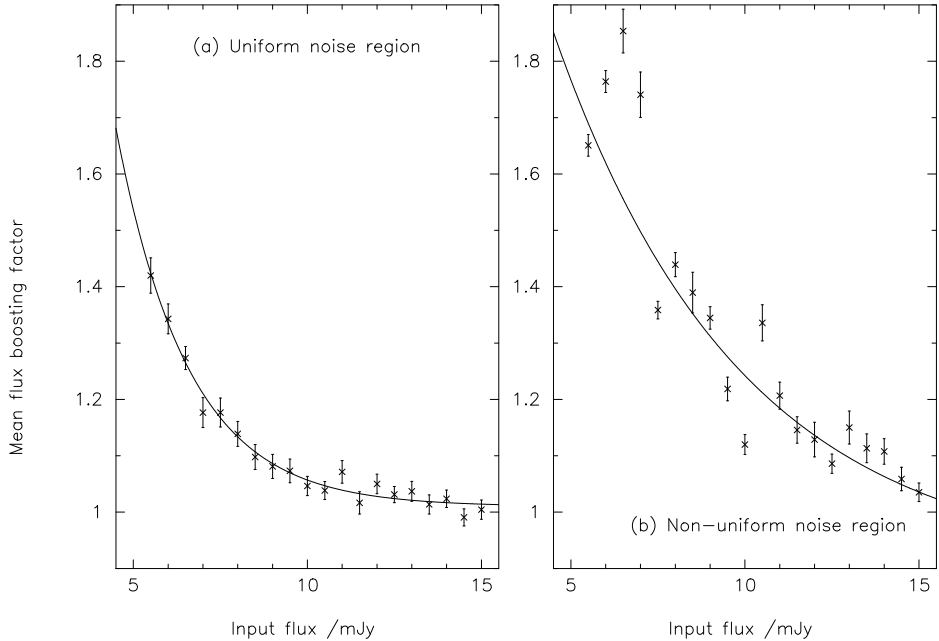}}
\caption{\small{The factor by which the input source flux is boosted
(when retrieved from the sub-mm map) plotted against input source flux density 
in the range $5-15$~mJy, based on a single template
source repeatedly being added into and retrieved ($>3.50\sigma$) from the real unconvolved
signal maps. The error bars give the standard error on the
mean. Plot (a) gives the results from the uniform noise region,
whereas plot (b) gives the results for the non-uniform edge regions
which have not received the full integration time. The solid curves show
a best-fit exponential-based model to the plotted data points.}
\label{fig:flux_boost}} 
\end{centering}
\end{figure*}

\begin{figure*}
\begin{centering}
\centerline{\epsfig{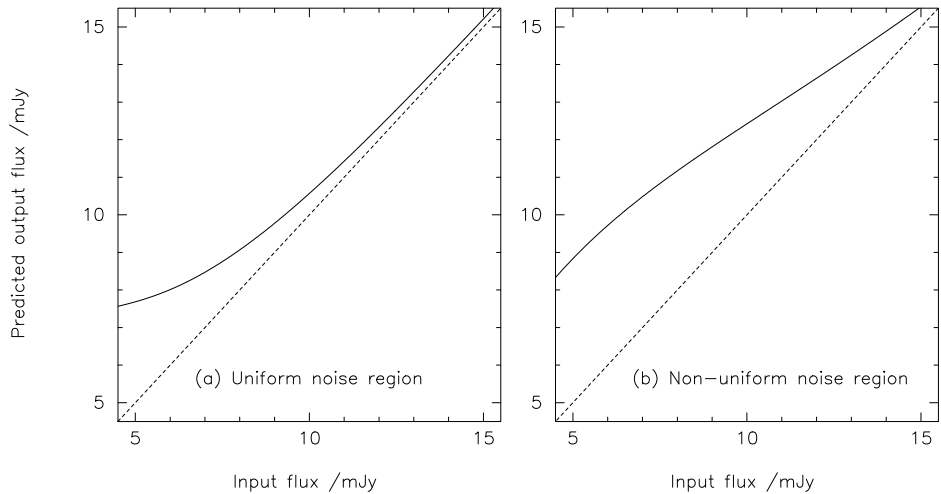}}
\caption{\small{Based on the fitted flux-density boosting curves 
shown in the preceding
Figure, these plots show the predicted output flux density against input
flux density in the range $5-15$~mJy. Plot (a) gives the results from the uniform noise region,
whereas plot (b) gives the results for the non-uniform edge regions
which have not received the full integration time. The dashed line in
both cases delineates the ideal case of
output flux density equal to input flux density.}
\label{fig:out_v_in}} 
\end{centering}
\end{figure*}

\section{Data reduction and map production}

The data have been reduced using two independent software packages;
SURF (Jenness et al. 1997) and an IDL based reduction package 
(Serjeant et al. 2001). 
The two mechanisms follow a very similar core
reduction process, the main difference being in the production of the final
maps. In addition to the output signal maps, the IDL routines 
automatically create corresponding uncorrelated noise maps by
considering the signal variance. The final images produced by the SURF and IDL
reductions were found to be consistent to within a few per cent of
each other. However, given the improved knowledge of the noise local
to each source, the IDL mechanism was adopted as the primary reduction
method and it is this process which we describe here.

After combining the positive and negative beams, the data were
flat-fielded and then corrected for atmospheric extinction. On the
nights when the CSO 225 GHz opacity meter made sufficiently good
measurements around the time of a map, each jiggle pointing of the
data was extinction-corrected by means of a polynomial fit to the
225~GHz opacity readings, interpolated to 850 and 450~${\rm \mu m}$ 
by means of the relations in Archibald et al. (2000b). 
If the CSO 225~GHz opacity meter was broken, or if the 225~GHz opacity 
measurements around the time of a jiggle map were poor, the extinction
correction was assessed instead by a linear interpolation between
consecutive skydips. 

An iterative procedure was employed to deglitch the data-sets and subtract any
residual sky emission not removed by chopping and nodding. Each
iteration made a time-dependent estimate of the noise for each
bolometer, removing any spikes by means of a $3\sigma$-clip. A
temporal modal sky level was determined by a fit to all bolometers in
the array and subtracted from the data. With each consecutive
iteration, the deglitching process makes a harder cut. Noisy
bolometers were assigned a low inverse variance weight in this way. 

To construct the final
images the signal data were binned into 1~arcsec pixels,
creating `zero-footprint' maps with a
corresponding noise value  determined from the signal
variance. The term `zero-footprint' is an analogy with the drizzling
algorithm (Fruchter \& Hook, 2002). A standard shift-and-add technique
takes the flux in a given detector pixel and places its flux into the
final map, over an area equivalent to one detector pixel projected on
the sky. Drizzling on the other hand takes the flux and and places it
into a smaller area in the final map. Simulations have shown that this
helps preserve information on small angular scales, provided that
there are enough observations to fill in the resulting gaps. The area
in the coadded map receiving the flux from one detector pixel is
termed the \emph{footprint}. Our method is an extreme example of
drizzling; we take the data from each 14.5 arcsecond bolometer and put
the flux into a very small footprint (the `zero-footprint'), one
arcsecond square. Unlike in the standard SURF reduction, there is no intrinsic smoothing or
interpolation between neighbouring pixels in this rebin
procedure. Although there is some degree of correlation between pixels
in the output zero-footprint \emph{signal} maps in terms of the beam
pattern, the corresponding pixel \emph{noise} values represent
individual measurements of the temporally varying sky noise averaged
over the dataset integration time, at a specific point on the sky, and
are hence statistically independent from their neighbours. In essence we have a very oversampled image with
statistically independent pixels.  Statistical non-independence would
refer to pixel-to-pixel crosstalk, which is not the case here.
A final
$4\sigma $-clip on the signal-to-noise was carried out to remove any remaining
`hot pixels'. A noise-weighted convolution with a beam-sized Gaussian
point spread function (PSF) produces realistic smoothed maps of the survey
areas and is able to account for variable signal-to-noise between
individual pixels.

Due to the overlap of individual jiggle-maps, and the fact that 
the observations were spread 
over 41 observing shifts, data missing due to bad bolometers are 
generally widely distributed, and only
one region of the final 450~${\rm \mu m}$ ELAIS N2 map has been noticeably
affected. This is in the region of RA 16:36:48.8 DEC +41:02:44, but is in 
fact a region which 
contains no significant 850~${\rm \mu m}$ sources. Noise arising from 
undersampling by bolometers at the edge of the
array has been evened out as far as possible by means of the tripod
positioning scheme and only significantly affects the perimeter of the
final mosaiced maps. Once these edges have been trimmed off, the
resulting images have almost uniform noise across the central
regions of the survey areas ($\sigma_{850}=2.5 \pm 0.7$~mJy/beam). 
The nature of the tripod positioning
scheme does, however, lead to a noisier border of thickness $\sim 1/2$ an
array width surrounding the uniform noise region due to incomplete observations
of one or more overlap positions, and hence less total integration time.

\subsection{Calibration}

Dunne \& Eales (2001) have
examined the issue of calibration errors in detail using the data from
the `SCUBA Local Universe Galaxy Survey' taken with the narrow band
850/450~${\rm \mu m}$ filters. They find that an
aperture calibration factor (ACF) is more accurate in determining
source flux densities than the method of determining the flux
conversion factor (FCF) by fitting
to the calibrator peaks. At 850~${\rm \mu m}$ the standard
deviation on the ACF is $\sim 6\%$, compared with $\sim 8\%$ when
applying an FCF fitted to the peak. The improvement is
particularly significant at 450~${\rm \mu m}$ where an aperture calibration
method yields a typical standard deviation of $\sim 10\%$ compared with
$\sim 18\%$ applying a flux conversion factor fitted to the peak. The greater
variation in gain when fitting to the calibrator peaks arises because of the
sensitivity of the peak flux density to changes in beam shape, due to sky
noise, pointing drifts, chop throw, and the dish shape (particularly
important at 450~${\rm \mu m}$). 

Our IDL pipeline does, however, have the considerable advantage of producing
uncorrelated noise maps, which have allowed us to apply a
maximum-likelihood method to measure the statistical significance of
each peak in our images, (see Section 4, and
Appendix A). Our source-extraction algorthim provides a
simultaneous fit to all peaks in the image, and is therefore able to
decouple any partially-confused sources whilst still recovering the
additional signal-to-noise yielded by the negative sidelobes. The
resulting best-fit model therefore allows us to measure the flux densities of
our significant detections ($>3.00 \sigma$) directly, as well as
yielding properly-quantified errors on this value. At 850~${\rm \mu m}$ the
accuracy gained by this approach more than compensates for the extra
$\sim 2\%$ error in the calibration and so we have chosen to employ
a calibration system based on fitting to the peak values. For
consistency, we have adopted the same approach when considering
the 450~${\rm \mu m}$ data \emph{in this paper}. 
The short-wavelength data will,
however, be considered in more detail by Fox et al. (2002).

In order to calibrate the data, each shift was divided in two and all
available information was utilised to calculate the mean 
flux-conversion factors applicable to
each half shift by fitting to the calibrator peaks. Uranus and Mars
were observed as primary calibrators when available, with the planetary
flux densities 
taken from the JCMT $\sc{FLUXES}$ program. Additionally, the secondary
calibrators CRL618, OH231.8, CRL2688 and IRC10216 were used. Since
IRC10216 is variable with a period of 635 days, it was assumed to have
constant flux density over an 8/9 night observing run, calculated using all
other calibration data for that run. Beam maps of the pointing sources
were used for tertiary calibration. A 30-arcsec chop-throw was used
for observing both the survey areas and the calibrators, in order to cancel out
the effects of beam distortion as far as possible. 
Taking into account the errors on
the calibrator flux densities, the variation in the 
flux-conversion factor over half a
shift, and errors in the measured atmospheric opacity, the average
calibration error over the survey maps was found to be 9\% at 
850~${\rm \mu m}$, and 20\% at 450~${\rm \mu m}$.

\subsection{Pointing Problems}

During the course of the survey, 3 different pointing problems are now
known to have afflicted the JCMT. 

First, prior to 28th July 1998, a
bug in the chopping software of SCUBA meant that the chop direction
was not updated once an
observation had begun, and consequently the negative sidelobes
accompanying a source in this data set will appear smeared and
therefore shallower than expected. However, the effect is small when
compared to the sky noise of $\simeq 2.5$~mJy/beam and affects only a small
portion of our data (51/502 maps).

Second, some of the jiggle-maps taken since June 1999 may have been
affected by an elevation drive pointing glitch (Coulson 2000). 
Observations appear to
suffer a 4 arcsec jump in the elevation when passing through
transit. Additionally, pointing on the opposite side of the meridian
to the survey field is likely to introduce a similar effect, of order
1.5--2 arcsec in magnitude. The fact that each area of sky is
covered by 12 individual jiggle-maps significantly reduces the
likelihood of a large increase in positional uncertainty and source
smearing. Under the most pessimistic assumptions, a maximum of 83/502 maps 
could be affected, of 
which a maximum of 3/12 cover any one area of sky.

Finally, data taken after 15th July 1999 were affected by a
non-synchronization of the GPS and SCUBA data acquisition computer
clocks. This problem has now been corrected for in the data reduction
(Jenness 2000).

\section{Source extraction technique}

The chopping-nodding mechanism of the telescope provides a valuable
method of discriminating between real detections and spurious noise
spikes in the data. The 30-arcsec chop throw is small enough to
fall onto the SCUBA array and so negative sidelobes, half the depth of
the peak flux, are produced on either side of a real source, 30
arcsec away from the position of the central peak. This side-lobe
signal can be recovered to boost the overall signal-to-noise ratio 
of a detection.

For well-separated sources, convolving
the images with the beam is formally the best method of source
extraction (Eales et al. 1999, 2000, Serjeant et al. 2001). However, following a careful
examination of the reduced data it became clear that some of the
sources were partially confused, particularly in the map of ELAIS N2
where the negative sidelobes of individual sources had overlapped and
were therefore somewhat deepened relative to both source
peaks. Consequently, in order to decouple any confused sidelobes a
source-extraction algorithm was devised based on a simultaneous 
maximum-likelihood fit to the flux densities of all potentially 
significant peaks in the beam-convolved maps. 

Using a peak-normalised beam map as a source template (CRL618 in the
case of the Lockman Hole East, or Uranus in the case of ELAIS N2), 
a basic model was constructed by centring a beam map at the
positions of all peaks $>3$~mJy as found
in the Gaussian-convolved maps. The heights of each of the
positioned beam maps were then calculated such that the
final multi-source model provided the best description of the sub-mm 
sky, as judged by a minimum $\chi^{2}$ calculation made feasible 
by the independent data-points and errors yielded by the zero-footprint 
IDL-reduced maps. A full mathematical description of this technique is 
provided in Appendix A.

The actual maps and source analysis are presented in Section 6. First, 
however, it is necessary to understand the selection effects that arise 
in an analysis of this sort.
 
\section{Simulations}

In order to assess the effects of confusion and noise on the
reliability of our source-extraction algorithm, Monte Carlo simulations
were carried out using the 220 out of 260 arcmin$^2$ of data taken by
the end of August 2000. The dependences of positional
error, completeness, and error in reclaimed flux density, on source
flux density and noise in the maps were determined by planting individual sources of
known flux density into the real ELAIS N2 and Lockman Hole East SCUBA
maps. This has the advantage of allowing us to test the 
source-reclamation process against the real noise properties of the
images. However, these simulations do not allow assessment of the
level to which false or confused sources can contaminate an extracted
source list. We have therefore also created a number of fully-simulated 
images of the survey areas by assuming a reasonable 
850~${\rm \mu m}$ source-counts model.
The results of analyzing these two sets of simulations are discussed in the
following two subsections.

\subsection{Simulations building on the real survey data}

\begin{figure*}
\begin{centering}
\centerline{\epsfig{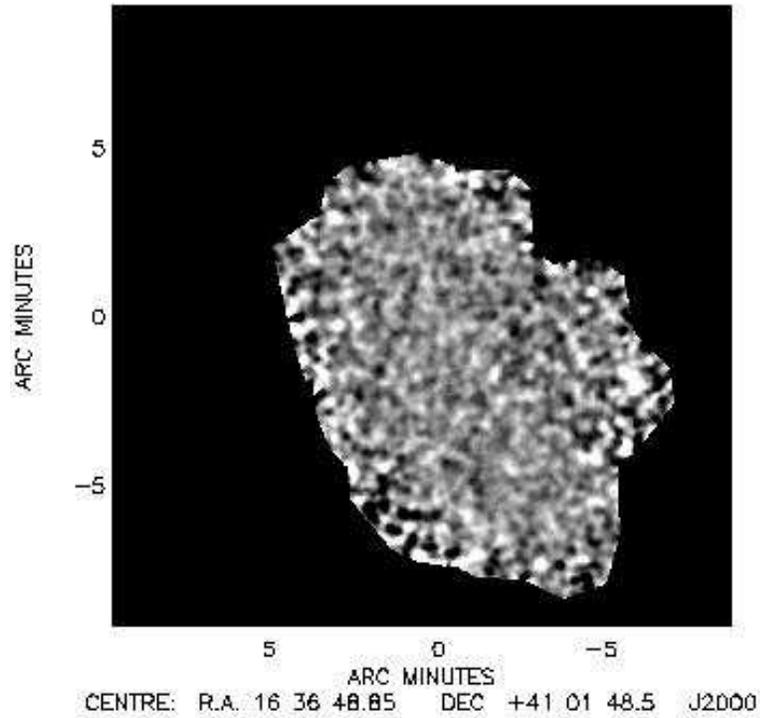}}
\caption{\small{A pure noise image of the ELAIS N2 region, created
from the real data by corrupting the astrometry of the individual
jiggle maps to produce a `scrambled' map. This image includes the
jiggle maps taken up to and including 16th August 2000. There are no
source detections at $>3.50\sigma$.}
\label{fig:scrambled_n2}} 
\end{centering}
\end{figure*}

\begin{figure*}
\begin{centering}
\centerline{\epsfig{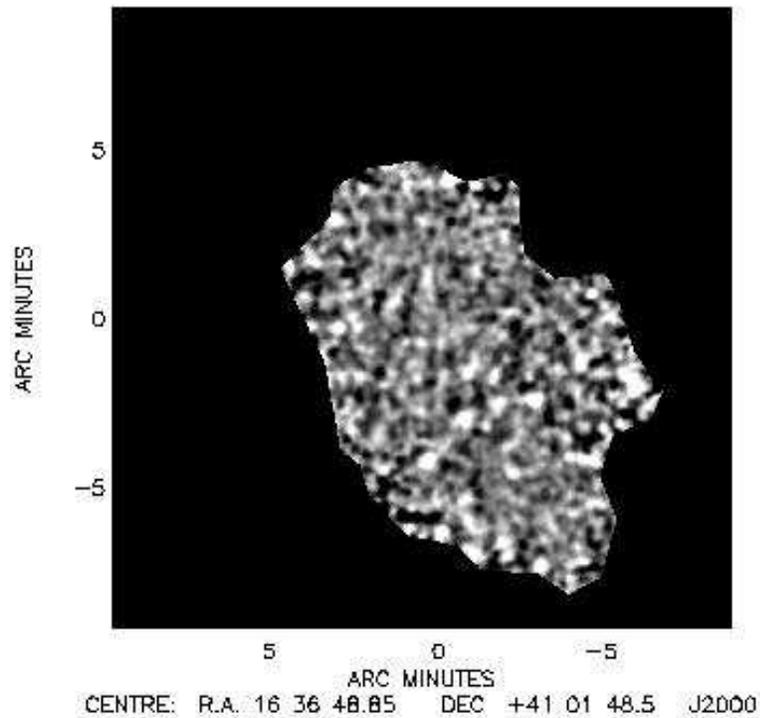}}
\caption{\small{Fully-simulated ELAIS N2 SCUBA 850~${\rm \mu m}$ map, 
using the adopted counts model given in Section 5.2. The image noise was simulated
using the scrambled map shown in Figure 5.}
\label{fig:simn2}} 
\end{centering}
\end{figure*}

\begin{figure*}
\begin{centering}
\centerline{\epsfig{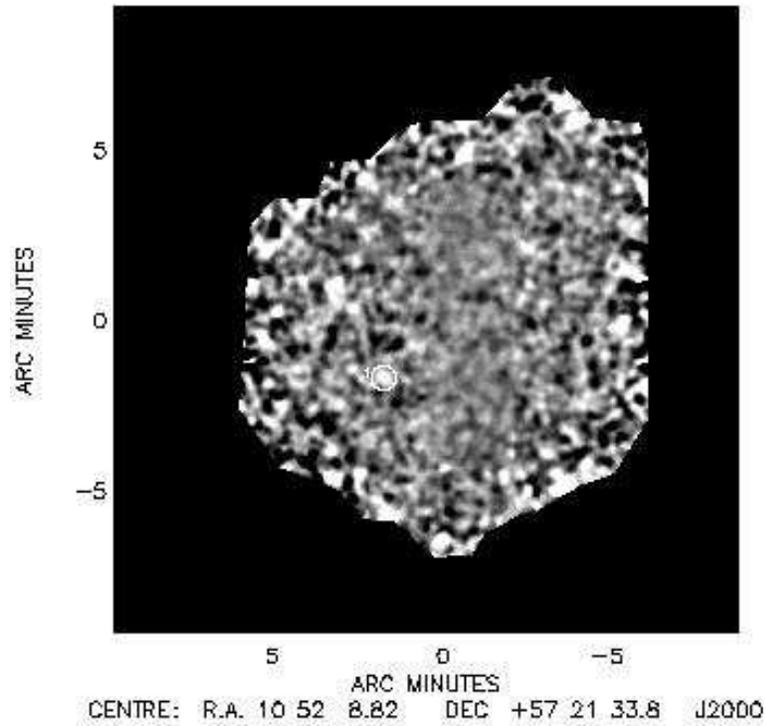}}
\caption{\small{A pure noise image of the Lockman Hole East region, created
from the real data by corrupting the astrometry of the individual
jiggle maps to produce a `scrambled' map. This image includes the
jiggle maps taken up to and including 16th August 2000. There is one
spurious source detection at $> 3.50 \sigma$ which is circled on the image.}
\label{fig:scrambled_lh}} 
\end{centering}
\end{figure*}

\begin{figure*}
\begin{centering}
\centerline{\epsfig{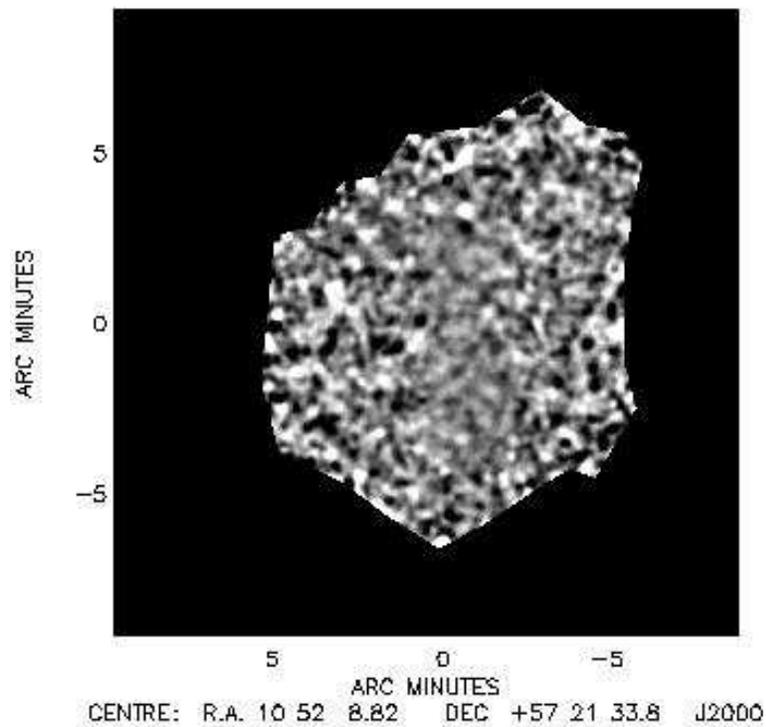}}
\caption{\small{Fully-simulated Lockman Hole East SCUBA 850~${\rm \mu m}$ map, 
using the adopted counts model given in Section 5.2. The image noise was
simulated using the scrambled map shown in Figure 7.}
\label{fig:simlh}} 
\end{centering}
\end{figure*}

A normalised beam map was used as a source
template. At 0.5~mJy flux intervals, from 5 to 15~mJy, individual sources
were added into the unconvolved zero-footprint signal maps. This was
done one fake source at a time, so as not to enhance significantly
any existing real confusion noise within the image. The
source-extraction algorithm was then re-run. This exercise was repeated for 100
different randomly-selected positions on each image, at each flux
level, so that source reclamation could be monitored as a function of
input flux and position/noise-level within the maps. Source
reclamation was deemed to have been successful if the 
source-extraction algorithm returned a source with signal-to-noise $> 3.50$
within less than half a beam width of the input position. When
analysing the output data, $\simeq 10\%$ of the input sources were
found to lie within half a beam width of another peak $> 5$~mJy
already present in the image. Any such source was excluded from
statistical analyses of positional error, completeness and error in
reclaimed flux density, since the proportion of blends produced 
by two sources, both
with comparably bright fluxes in the region of 8~mJy, should be
negligible within our sample. 

Figs 1 to 4 show the dependences of mean positional
error, percentage of sources returned, systematic offset in returned 
flux density, and predicted output against input flux density, 
for both regions of uniform and non-uniform noise. The error bars show
the standard error on the mean. The solid lines in Fig 1, and solid
curves in Figs 2 and 3, are best-fit models to the plotted data
points. From Fig 3, we
see that the effect of noise and confusion is to produce systematic 
`flux boosting', the
retrieved flux density always being greater than the input value. The
solid curves in Fig 4 show the predicted output versus input
flux density based on the fitted flux-density boosting models of Fig 3. 
We find that there is a strong correlation
with input flux density for each of these quantities, the scatter being
somewhat larger in the areas of non-uniform noise lying towards the
edge, as would be expected. The positional error and flux-density boosting
factors decrease with increasing input flux density, whilst the 
percentage of
sources returned at $>3.50 \sigma$ increases.
The effect of additional noise 
in the non-uniform regions considerably increases the
uncertainty in positional error, as well as the mean flux-density boosting
factor. It also serves to typically half the percentage of sources
reclaimed. A summary of the results at 5, 8, 11 and 14~mJy is given
in Table 1. The integral completeness down to each flux-density 
limit has been calculated by
a counts-slope-weighted sum of the fraction of sources returned at
each specific flux-density level, based on an adopted 850~${\rm \mu m}$
source counts model consistent with the measured source counts, given in
Section 5.2. As detailed in Table 1, the typical completeness
at 8~mJy is $\simeq 80 \% $ within the uniform noise regions of the
images, with flux-density boosting factors of $\simeq 15 \% $. By 11~mJy
completeness has risen to $\simeq 90 \% $, while flux-density boosting has
dropped to $< 10 \% $.

\clearpage

\begin{table*}
\begin{tabular}{|l|c|c|c|c|c|c|} \hline
 & ELAIS N2  & ELAIS N2 & Lockman Hole E & Lockman Hole E &
 All & All\\
 & Uniform noise & Full area & Uniform noise & Full area & Uniform
 noise & Full area \\ \hline
Area (sq. arcmin) & 63 & 102 & 55 & 122 & 118 & 224\\ \hline
Sources $>8$~mJy in & 16 & 28 & 12 & 18 & 28 & 46 \\ 
Sources $>8$~mJy out & 18 & 25 & 16 & 20 & 34 & 45 \\
Real sources $>8$~mJy out & 12 & 16 & 10 & 13 & 22 & 29 \\
Confused sources $>8$~mJy & \phantom{1}1 & \phantom{0}1 & \phantom{0}2 & \phantom{0}2 & \phantom{0}3 & \phantom{0}3  \\
Real boosted sources & \phantom{1}5 & \phantom{0}8 & \phantom{0}4 & \phantom{0}5 & \phantom{0}9 & 13 \\
$>5$~mJy in, $>8$~mJy out &  &  &  &  &  &  \\ \hline
\% completeness & $75.0 \pm 21.7$ & $\phantom{1}57.1 \pm 14.3$ & $83.3 \pm 26.4$ & $72.2
 \pm 20.0$ & $78.6 \pm 16.8$ & $\phantom{1}63.0 \pm 11.7$ \\
\% confused/random & $\phantom{0}5.6 \pm \phantom{0}5.6$ & $\phantom{10}4.0 
\pm \phantom{0}4.0$ & $12.5 \pm \phantom{0}8.8$ & $10.0
 \pm \phantom{0}7.1$ & $\phantom{0}8.8 \pm \phantom{0}5.1$ & $\phantom{10}6.7 
\pm \phantom{0}3.8$ \\
\% count correction factor & $88.9 \pm 22.2$ & $112.0 \pm 21.2$ & $75.0 \pm
 21.7$ & $90.0 \pm 21.2$ & $82.4 \pm 15.6$ & $102.2 \pm 15.1$ \\ \hline
\end{tabular}
\label{table:compsim1}\caption{\small Results of the completeness
 simulations at the $4.00\sigma$ significance level, in which
 each field was fully simulated 4 times using an adopted counts
 model consistent with the measured source counts. The sources were then retrieved by means of the
 maximum-likelihood estimator outlined in Section 4.}
\end{table*}

\begin{table*}
\begin{tabular}{|l|c|c|c|c|c|c|} \hline
 & ELAIS N2  & ELAIS N2 & Lockman Hole E & Lockman Hole E &
 All & All\\
 & Uniform noise & Full area & Uniform noise & Full area & Uniform
 noise & Full area \\ \hline
Area (sq. arcmin) & 63 & 102 & 55 & 122 & 118 & 224\\ \hline
Sources $>8$~mJy in & 16 & 28 & 12 & 18 & 28 & 46 \\ 
Sources $>8$~mJy out & 23 & 32 & 19 & 24 & 42 & 56  \\ 
Real sources $>8$~mJy out & 13 & 17 & 11 & 14 & 24 & 31 \\ 
Confused sources $>8$~mJy & \phantom{0}4 & \phantom{0}5 & \phantom{0}4 & \phantom{0}5 & \phantom{0}8 & 10  \\ 
Real boosted sources & \phantom{0}6 & 10 & \phantom{0}4 & \phantom{0}5 & 10 & 15 \\ 
$>5$~mJy in, $>8$~mJy out&  &  &  &  &  &  \\ \hline
\% completeness & $81.3 \pm 22.5$ & $60.7 \pm 14.7$  & $91.7 \pm 27.6$
 & $77.8 \pm 20.8$ & $85.7 \pm 17.5$  & $67.4 \pm 12.1$ \\ 
\% confused/random &  $17.4 \pm \phantom{0}8.7$ & $15.6 \pm \phantom{0}7.0$  
& $21.1 \pm
 10.5$ & $20.8 \pm \phantom{0}9.3$  & $19.0 \pm \phantom{0}6.7$ & 
$17.9 \pm \phantom{0}5.6$  \\ 
\% count correction factor& $69.6 \pm 17.4$ & $87.5 \pm 16.5$  &
 $63.0 \pm 18.2$  & $75.0 \pm 17.7$  & $66.7 \pm 12.6$  & $82.1 \pm 12.1$ \\ \hline
\end{tabular}
\label{table:compsim2}\caption{\small Results of the completeness
 simulations at the $3.50\sigma$ significance level, in which
 each field was fully simulated 4 times using an adopted counts
 model consistent with the measured source counts. The sources were then retrieved by means of the
 maximum-likelihood estimator outlined in Section 4.}
\end{table*}

\begin{table*}
\begin{tabular}{|l|c|c|c|c|c|c|} \hline
 & ELAIS N2  & ELAIS N2 & Lockman Hole E & Lockman Hole E &
 All & All\\
 & Uniform noise & Full area & Uniform noise & Full area & Uniform
 noise & Full area \\ \hline
Area (sq. arcmin) & 63 & 102 & 55 & 122 & 118 & 224\\ \hline
Sources $>8$~mJy in & 16 & 28 & 12 & 18 & 28 & 46 \\ 
Sources $>8$~mJy out & 26 & 50 & 31 & 49 & 57 & 99  \\ 
Real sources $>8$~mJy out & 13 & 19 & 11 & 14 & 24 & 33 \\ 
Confused sources $>8$~mJy & \phantom{0}7 & 20 & 16 & 28 & 23 & 48  \\ 
Real boosted sources & \phantom{0}6 & 11 & \phantom{0}4 & \phantom{0}7 & 
10 & 18 \\ 
$>5$~mJy in, $>8$~mJy out&  &  &  &  &  &  \\ \hline
\% completeness & $81.3 \pm 22.5$ & $67.9 \pm 15.6$  & $91.7 \pm 27.6$
 & $77.8 \pm 20.8$ & $85.7 \pm 17.5$  & $71.7 \pm 12.5$ \\ 
\% confused/random &  $26.9 \pm 10.2$ & $40.0 \pm \phantom{0}8.9$  & $51.6 \pm
 12.9$ & $57.1 \pm 10.8$  & $40.4 \pm \phantom{0}8.4$ & $48.5 \pm \phantom{0}7.0$  \\ 
\% count correction factor & $61.5 \pm 15.4$ & $56.0 \pm 10.6$  &
 $38.7 \pm 11.2$  & $36.7 \pm \phantom{0}8.7$  & $49.1 \pm \phantom{0}9.3$  & $46.5 \pm \phantom{0}6.9$ \\ \hline
\end{tabular}
\label{table:compsim3}\caption{\small Results of the completeness
 simulations at the $3.00\sigma$ significance level, in which
 each field was fully simulated 4 times using an adopted counts
 model consistent with the measured source counts. The sources were then retrieved by means of the
 maximum-likelihood estimator outlined in Section 4.}
\end{table*}

\clearpage

\subsection{Completely simulated maps}

Adopting an 850~${\rm \mu m}$ source counts model consistent with the
measured point source counts, and which does not over-predict the FIR
background, we have created fully-simulated versions of our two survey
fields. In order to produce a realistic model of the background
counts, the number of sources expected at 0.1~mJy intervals
was Gaussian randomised and then the appropriate sources 
were planted into the
simulated data sets at random positions. The whole image was then
convolved with the beam. At all flux-density levels the sources were assumed
to be un-clustered (although see Section 9.4 and Almaini et al. 2002).

We then constructed a pure noise image for each of our two fields 
by scrambling the astrometry
of the individual SCUBA observations used to produce the maps. These
may be seen in Figs 5 and 7. After
creating the scrambled noise maps we first ran the source-extraction
algorithm on these to assess the frequency with which spurious sources
would artificially appear on the basis of random statistics. The noise
images yielded only a single artificial source at $>3.50 \sigma$, and
a total of 12 spurious sources at $> 3.00 \sigma$, which is not unexpected
given the $\simeq 1200$ beams across the areas of these maps. Adding
the simulated background counts to the scrambled maps produces the
final simulated images. The source-extraction algorithm was applied as
previously.

Four realisations of the sky in each field were generated for the
source counts model 
\be \frac{dN(>S)}{dS} = \frac{N_{0}}{S_{0}} \left(
 \left(\frac{S}{S_{0}} \right)^{\alpha} + \left(\frac{S}{S_{0}}
\right)^{\beta} \right)^{-1} \ee 
where $N_{0} = 1.1 \times 10^{4} \mathrm{deg^{-2}mJy^{-1}}$ , $S_{0} =
1.79$~mJy, $\alpha~=~0.0$, and $\beta~=~3.13$. Examples of the simulated
ELAIS N2 and Lockman Hole East maps for our adopted counts
model may be seen in Figs 6 and 8. 

Tables 2 to 4 summarise the results at 4.00, 3.50 and
3.00$\sigma$ respectively for the simulated fields at 8~mJy. 
In these tables and the
discussion that follows, we have adopted the practical criterion that
any source retrieved at the 8~mJy level which ultimately could not be
identified with a single source brighter than 5~mJy (and therefore
could not be confirmed through interferometric follow-up for example),
should be regarded as spurious or the result of confusion. In the
uniform noise regions, contamination from spurious sources or confused
sources $< 5$~mJy is $\simeq 40\%$ at S/N~$> 3.00$, 
dropping to $\simeq 20\%$ at
S/N~$> 3.50$ and $<10\%$ at S/N~$> 4.00$. The completeness
at a measured flux density of 8~mJy is $\simeq 85\%$ within the regions of
uniform noise, consistent with the completeness determined from the
simulations using the real survey data. We also find that although the
real source counts (ie. total number of sources out minus the number of
spurious sources) at $> 3.50\sigma$ agree well with the number planted
into the simulations at 8~mJy, this is partly by chance; the number of
sources boosted from a lower flux-density level, between 5 and 8~mJy, is
approximately the same as the number of real sources $> 8$~mJy which
are not retrieved by the source-extraction algorithm. By lowering our
signal-to-noise threshold from 3.50 to 3.00, there is very little
improvement (if any) in completeness, however the proportion of
spurious sources detected is much more serious. These results provide
a clear motivation for setting a signal-to-noise threshold of $>3.50$ when
determining the source number counts as opposed to the $>3.00\sigma$ level
applied in some other surveys. Our results are consistent with those
of Hughes \& Gazta\~naga (2000), who have simulated sub-millimetre
galaxy surveys for a range of wavelengths, spatial resolutions and
flux densities. 

\begin{figure*}
\vspace*{19cm}
\includegraphics{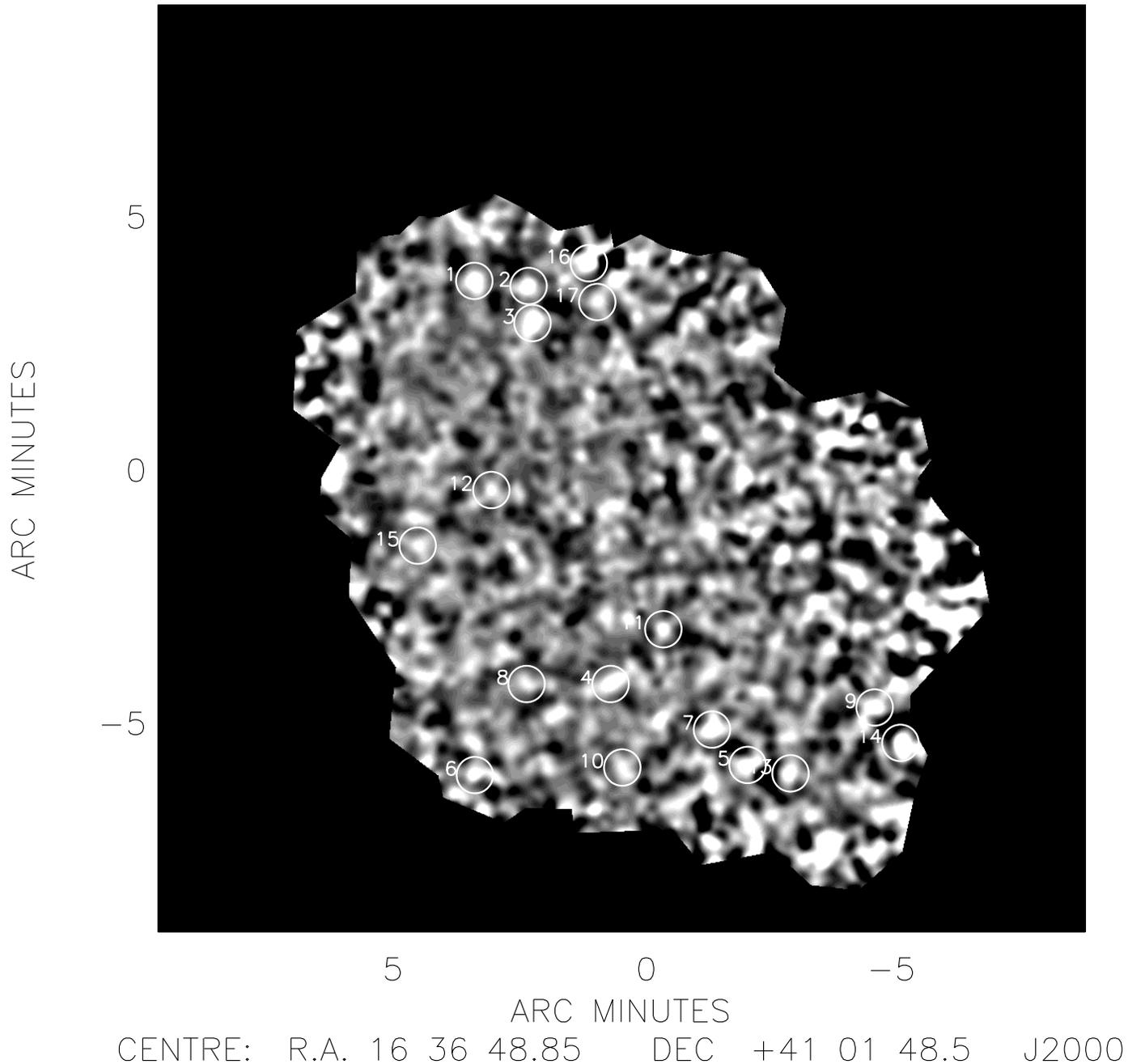}
\caption{\small{The 850~${\rm \mu m}$ image of the ELAIS N2 field, 
smoothed with a beam-size Gaussian (14.5 arcsec FWHM). The
numbered circles highlight those sources found at a significance of
$>3.50$, the labels corresponding to the numbers in Table 5.}
\label{fig:elaisn2_850}} 
\end{figure*}

\begin{table*}
\begin{tabular}{|l|c|c|r|c|c|l|c|c|c|c|} \hline
Source & RA & DEC & $S_{850}\phantom{00}$ & S/N       & Noise & $S_{450}$ or $3\sigma$    & S/N        & $S_{450}$ of any& S/N   & Distance \\
        &(J2000)&(J2000)&    /mJy\phantom{00}    & 850~${\rm \mu m}$& 
Region
& up. lim. at& 450~${\rm \mu m}$ & $>3\sigma$ source& 450~${\rm \mu m}$ & from\\
       &    &     &           &           &              & 850~${\rm \mu m}$ pos.  &  at & $< 8.0^{\prime \prime}$ from   & at  & 850~${\rm \mu m}$     \\ 
       &    &     &           &           &              & /mJy          &  
850~${\rm \mu m}$        & 850~${\rm \mu m}$ pos & 450~${\rm \mu m}$ & peak \\ 
 & & & & & & & pos & /mJy & pos & /arcsec\\ \hline

01 & 16:37:04.3 & 41:05:30 &  $11.2 \pm 1.6$ &  8.59 & uniform & $\phantom{<}30 \pm 10$ & 3.82 & $32 \pm 10$ & 4.08 & 2.0 \\
02 & 16:36:58.7 & 41:05:24 &  $10.7 \pm 2.0$ &  6.27 & uniform & $\phantom{<}30 \pm 9$ & 4.10 & $30 \pm \phantom{1}9$ & 4.16 & 2.0 \\
03 & 16:36:58.2 & 41:04:42 &  $ 8.5 \pm 1.6$ &  5.86 & uniform & $<26$ & \\
04 & 16:36:50.0 & 40:57:33 &  $ 8.2 \pm 1.7$ &  5.18 & uniform & $<34$ & \\
05 & 16:36:35.6 & 40:55:58 &  $ 8.5 \pm 2.2$ &  4.16 & uniform & $<35$ & \\
06 & 16:37:04.2 & 40:55:45 &  $ 9.2 \pm 2.4$ &  4.13 & uniform & $<91$
& & $73 \pm 28$ & 3.09 & 4.5 \\
07 & 16:36:39.4 & 40:56:38 &  $ 9.0 \pm 2.4$ &  4.07 & uniform & $<39$ & \\ \hline
08 & 16:36:58.8 & 40:57:33 &  $ 5.1 \pm 1.4$ &  3.82 & uniform & $<43$ \\
09 & 16:36:22.4 & 40:57:05 &  $ 9.0 \pm 2.5$ &  3.76 & non-uni & $<88$ \\
10 & 16:36:48.8 & 40:55:54 &  $ 5.4 \pm 1.5$ &  3.69 & uniform & $<47$
& &  $39 \pm 15$ & 3.18 & 5.7 \\
11 & 16:36:44.5 & 40:58:38 &  $ 7.1 \pm 2.0$ &  3.67 & uniform & $<48$ \\
12 & 16:37:02.5 & 41:01:23 &  $ 5.5 \pm 1.6$ &  3.65 & uniform & $<42$ \\
13 & 16:36:31.2 & 40:55:47 &  $ 6.3 \pm 1.9$ &  3.56 & uniform & $<68$ \\
14 & 16:36:19.7 & 40:56:23 &  $11.2 \pm 3.3$ &  3.55 & non-uni & $<43$ \\
15 & 16:37:10.2 & 41:00:17 &  $ 5.0 \pm 1.5$ &  3.52 & uniform & $<18$ \\
16 & 16:36:52.3 & 41:05:52 &  $12.9 \pm 3.9$ &  3.51 & non-uni & $<18$ \\
17 & 16:36:51.4 & 41:05:06 &  $ 5.7 \pm 1.7$ &  3.50 & uniform & $<23$
\\ \hline \hline
18 & 16:36:11.4 & 40:59:26 &  $20.8 \pm 6.2$ &  3.49 & non-uni & $<138$ \\
19 & 16:36:35.9 & 41:01:38 &  $ 9.4 \pm 2.8$ &  3.49 & uniform & $<76$ \\
20 & 16:36:49.4 & 41:04:17 &  $ 6.6 \pm 2.0$ &  3.48 & uniform & $\phantom{<}22 \pm 8$ & 3.32 & $20 \pm \phantom{1}8$ & 3.05 & 3.2 \\
21 & 16:37:10.5 & 41:00:49 &  $ 4.9 \pm 1.5$ &  3.47 & uniform & $<23$ \\
22 & 16:36:27.9 & 40:54:04 &  $13.4 \pm 4.1$ &  3.46 & non-uni & $<102$ \\
23 & 16:37:19.5 & 41:01:38 &  $12.4 \pm 3.8$ &  3.45 & non-uni & $<40$ \\
24 & 16:36:26.9 & 41:02:23 &  $10.4 \pm 3.2$ &  3.43 & non-uni & $<44$ \\
25 & 16:36:18.3 & 40:59:12 &  $12.1 \pm 3.8$ &  3.37 & non-uni & $<57$ \\
26 & 16:36:32.0 & 41:00:05 &  $ 9.7 \pm 3.1$ &  3.26 & uniform & $<27$ \\
27 & 16:36:48.3 & 41:03:52 &  $ 6.6 \pm 2.1$ &  3.25 & uniform & $<21$ \\
28 & 16:36:47.2 & 41:04:48 &  $ 6.3 \pm 2.0$ &  3.24 & uniform & $<22$ \\
29 & 16:36:24.1 & 40:59:35 &  $ 9.9 \pm 3.3$ &  3.14 & uniform & $<70$ \\
30 & 16:37:07.5 & 41:02:37 &  $ 4.6 \pm 1.5$ &  3.13 & uniform & $<40$ \\
31 & 16:36:28.1 & 41:01:41 &  $ 6.8 \pm 2.3$ &  3.07 & uniform & $<36$ \\
32 & 16:36:39.8 & 41:00:34 &  $ 5.4 \pm 1.8$ &  3.07 & uniform & $<27$ \\
33 & 16:36:50.5 & 40:58:54 &  $ 4.8 \pm 1.6$ &  3.06 & uniform & $<34$ \\
34 & 16:36:27.0 & 40:58:15 &  $ 6.9 \pm 2.3$ &  3.05 & uniform & $<36$ \\
35 & 16:36:44.8 & 40:56:51 &  $ 5.5 \pm 1.9$ &  3.02 & uniform & $<23$ \\
36 & 16:36:52.9 & 41:02:52 &  $ 3.8 \pm 1.3$ &  3.00 & uniform & n/a \\ \hline
\end{tabular}
\label{table:n2sources}\caption{\small Sources retrieved in the
long-wavelength ELAIS N2 image above the $3.00\sigma$ significance level
using the source-extraction algorithm outlined in Section 4. We used a
peak-normalised calibrator beam-map as a source template, and
constructed a basic model by centring a beam-map at the positions of
all peaks $>3$~mJy as found in the 14.5$^{\prime \prime}$ 
FWHM Gaussian convolved
maps. The heights of each of the positioned beam maps were allowed to
vary simultaneously such that the final model satisfied a minimised
$\chi^{2}$ fit. We suggest that follow-up observations be confined to
those detections $>3.50\sigma$ given the serious problem of
contamination by spurious sources at the $3.00\sigma$
level. Calibration errors of 9\% and 20\% at 850~${\rm \mu m}$ and 450~${\rm 
\mu m}$ respectively, are included in the error on the flux densities.}
\end{table*}

\begin{figure*}
\vspace*{19cm}
\includegraphics{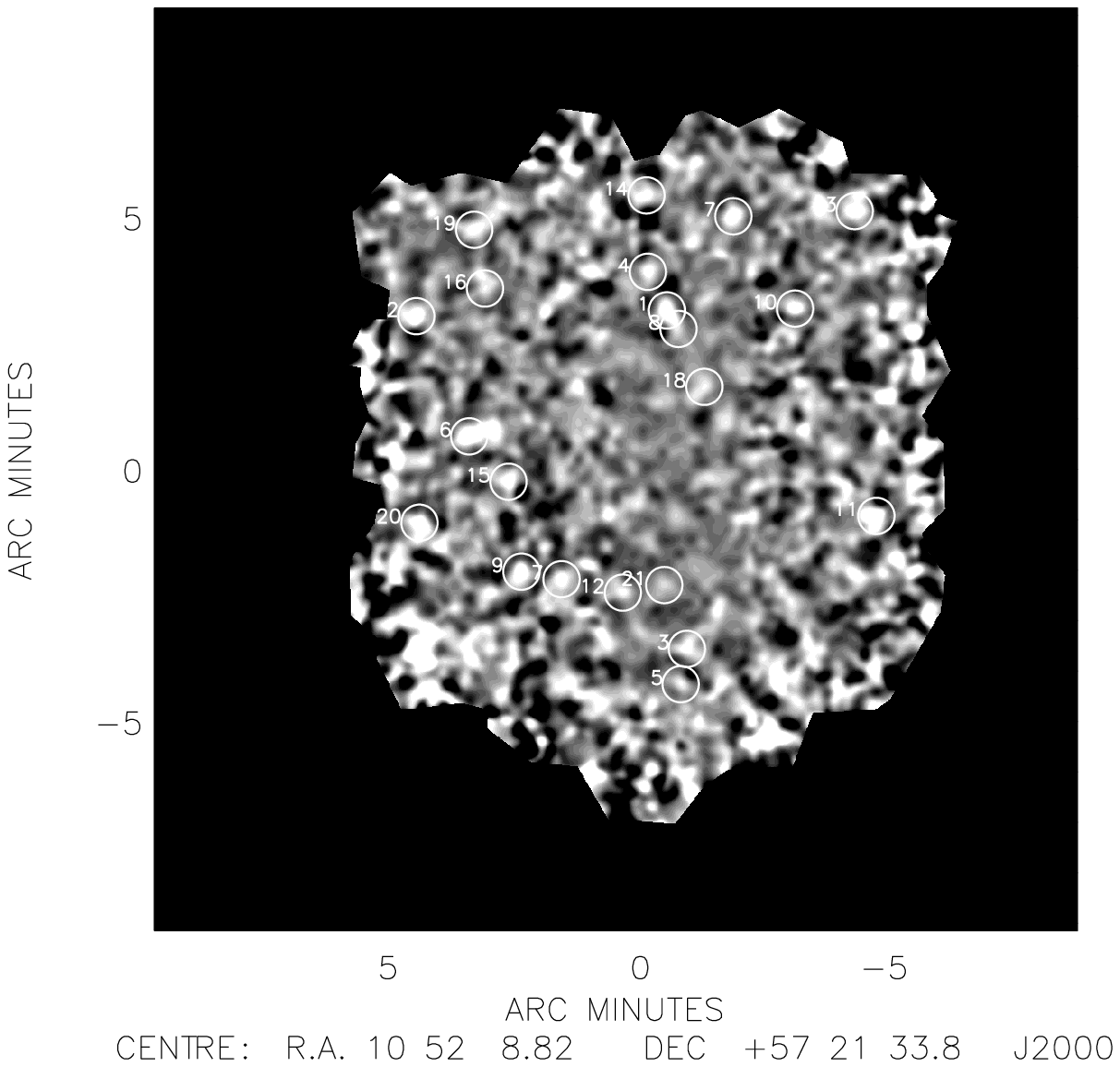}
\caption{\small{The 850~${\rm \mu m}$ image of the Lockman Hole East field,
smoothed with a beam-size Gaussian (14.5 arcsec FWHM). The
numbered circles highlight those sources found at a significance of
$>3.50$, the labels corresponding to the numbers in Table 6.}
\label{fig:lockman_850}} 
\end{figure*}

\begin{table*}
\begin{tabular}{|l|c|c|r|c|c|l|c|c|c|c|} \hline

Source & RA & DEC & $S_{850}\phantom{00}$ & S/N       & Noise & $S_{450}$ or $3\sigma$    & S/N        & $S_{450}$ of any& S/N   & Distance \\
        &(J2000)&(J2000)&    /mJy\phantom{00}    & 850~${\rm \mu m}$& 
Region
& up. lim. at& 450~${\rm \mu m}$ & $>3\sigma$ source& 450~${\rm \mu m}$ & from\\
       &    &     &           &           &              & 850~${\rm \mu m}$ pos.  &  at & $< 8.0^{\prime \prime}$ from   & at  & 850~${\rm \mu m}$     \\ 
       &    &     &           &           &              & /mJy          &  
850~${\rm \mu m}$        & 850~${\rm \mu m}$ pos & 450~${\rm \mu m}$ & peak \\ 
 & & & & & & & pos & /mJy & pos & /arcsec \\ \hline

01 & 10:52:01.4 & 57:24:43 &  $10.5 \pm 1.6$ &  8.10 & deep  & $<\phantom{1}43$ & & $45 \pm 13$ & 4.64 & 5.0 \\
02 & 10:52:38.2 & 57:24:36 &  $10.9 \pm 2.4$ & 5.22 & uniform & $<\phantom{1}41$ & &  &  &  \\

03 & 10:51:58.3 & 57:18:01 &  $ 7.7 \pm 1.7$ &  5.06 & deep  & $<\phantom{1}54$ & & $39 \pm 14$ & 3.38 & 3.2 \\
04 & 10:52:04.1 & 57:25:28 &  $ 8.3 \pm 1.8$ &  5.03 & deep  & $<\phantom{1}51$ \\
05 & 10:51:59.3 & 57:17:18 &  $ 8.6 \pm 2.0$ &  4.57 & deep  & $<\phantom{1}34$ \\
06 & 10:52:30.6 & 57:22:12 &  $11.0 \pm 2.6$ &  4.50 & uniform & $<101$ \\
07 & 10:51:51.5 & 57:26:35 &  $ 8.1 \pm 1.9$ &  4.50 & deep  & $<\phantom{1}86$ \\
08 & 10:52:00.0 & 57:24:21 &  $ 5.1 \pm 1.3$ &  4.38 & deep  & $<\phantom{1}17$ \\
09 & 10:52:22.7 & 57:19:32 &  $12.6 \pm 3.2$ &  4.20 & uniform & $<\phantom{1}47$ \\
10 & 10:51:42.4 & 57:24:45 &  $12.2 \pm 3.1$ &  4.18 & uniform & $<124$ \\
11 & 10:51:30.6 & 57:20:38 &  $13.5 \pm 3.5$ &  4.09 & non-uni & $<\phantom{1}90$ &  & $65 \pm 24$ & 3.32 & 6.0  \\
12 & 10:52:07.7 & 57:19:07 &  $ 6.2 \pm 1.6$ &  4.01 & deep  & $<\phantom{1}55$ \\ \hline
13 & 10:51:33.6 & 57:26:41 &  $ 9.8 \pm 2.8$ &  3.69 & uniform & $<\phantom{1}90$ \\
14 & 10:52:04.3 & 57:26:59 &  $ 9.5 \pm 2.8$ &  3.61 & uniform & $<274$ \\
15 & 10:52:24.6 & 57:21:19 &  $11.7 \pm 3.4$ &  3.60 & uniform & $<\phantom{1}53$ \\
16 & 10:52:27.1 & 57:25:16 &  $ 6.1 \pm 1.8$ &  3.56 & uniform & $<\phantom{1}39$ \\
17 & 10:52:16.8 & 57:19:23 &  $ 9.2 \pm 2.7$ &  3.55 & uniform & $<\phantom{1}57$ \\
18 & 10:51:55.7 & 57:23:12 &  $ 4.5 \pm 1.3$ &  3.55 & deep  & $<\phantom{1}36$ \\
19 & 10:52:29.7 & 57:26:19 &  $ 5.5 \pm 1.6$ &  3.54 & uniform & $<\phantom{1}39$ \\
20 & 10:52:37.7 & 57:20:30 &  $10.3 \pm 3.1$ &  3.51 & non-uni & $<\phantom{1}73$ \\
21 & 10:52:01.7 & 57:19:16 &  $ 4.5 \pm 1.3$ &  3.50 & deep  &
$<\phantom{1}35$ \\ \hline \hline
22 & 10:52:05.7 & 57:20:53 &  $ 4.7 \pm 1.4$ &  3.49 & deep  & $<\phantom{1}52$ \\
23 & 10:51:47.0 & 57:24:51 &  $ 7.4 \pm 2.2$ &  3.48 & uniform & $<\phantom{1}60$ \\
24 & 10:51:42.9 & 57:24:12 &  $11.4 \pm 3.4$ &  3.47 & uniform & $<\phantom{1}63$ \\
25 & 10:52:36.0 & 57:18:20 &  $12.7 \pm 3.8$ &  3.46 & non-uni & $<137$ \\
26 & 10:52:27.3 & 57:19:06 &  $ 8.3 \pm 2.6$ &  3.39 & uniform & $<\phantom{1}59$ \\
27 & 10:51:53.8 & 57:18:47 &  $ 5.5 \pm 1.7$ &  3.38 & deep  & $<\phantom{1}33$ \\
28 & 10:52:34.6 & 57:20:02 &  $10.2 \pm 3.2$ &  3.31 & uniform & $<\phantom{1}74$ \\
29 & 10:52:16.4 & 57:25:07 &  $ 6.7 \pm 2.1$ &  3.30 & uniform & $<\phantom{1}28$ \\
30 & 10:52:42.2 & 57:18:28 &  $11.2 \pm 3.6$ &  3.25 & non-uni & $<\phantom{1}92$ \\
31 & 10:52:03.9 & 57:20:07 &  $ 4.0 \pm 1.3$ &  3.24 & deep  & $<\phantom{1}22$ \\
32 & 10:52:00.0 & 57:20:39 &  $ 4.3 \pm 1.4$ &  3.22 & deep  & $<\phantom{1}24$ \\
33 & 10:51:33.8 & 57:19:29 &  $ 8.1 \pm 2.6$ &  3.20 & uniform & $<\phantom{1}75$ \\
34 & 10:52:09.9 & 57:20:40 &  $ 8.5 \pm 2.8$ &  3.16 & uniform & $<\phantom{1}77$ \\
35 & 10:51:57.6 & 57:26:03 &  $ 6.7 \pm 2.3$ &  3.02 & uniform & $<\phantom{1}91$ \\
36 & 10:52:03.5 & 57:16:54 &  $ 8.0 \pm 2.8$ &  3.00 & uniform & $<\phantom{1}29$ \\ \hline
\end{tabular}
\label{table:lhsources}\caption{\small Sources retrieved in the
Lockman Hole East field above the $3.00\sigma$ significance level using
the source extraction algorithm outlined in Section 4.We used a
peak-normalised calibrator beam-map as a source template, and
constructed a basic model by centring a beam-map at the positions of
all peaks $>3$~mJy as found in the 14.5'' FWHM Gaussian convolved
maps. The heights of each of the positioned beam maps were allowed to
vary simultaneously such that the final model satisfied a minimised
$\chi^{2}$ fit. We suggest that follow-up observations are confined to
those detections $>3.50\sigma$ given the serious problem of
contamination by spurious sources at the $3.00\sigma$ level.Calibration errors of 9\% and 20\% at 850~${\rm \mu m}$ and 450~${\rm \mu
m}$ respectively, are included in the error on the flux densities.}
\end{table*}

\section{Sources}

Sources down to a significance level of $> 3.00\sigma$ at 850~${\rm \mu m}$
are listed in Table 5 for ELAIS N2 and Table 6 for the Lockman
Hole East, ranked in order of formal significance as determined from
our maximum-likelihood extraction technique. The top section of each
table, containing 7 and 12 sources in ELAIS N2 and Lockman Hole East
respectively, is confined to sources for which the S/N $>4.00$, and
the middle section to sources for which S/N $>3.50$. Since our
simulations indicate that 70-80\% of these sources are expected to be
real individual sources with flux densities within $\simeq25\%$ of the
extracted flux density, we recommend that follow-up observations are
confined to these abbreviated S/N $>3.50$ source lists. The 
850~${\rm \mu m}$
flux densities in column 4 are the best-fit scaling factors
determined by our extraction algorithm, with the quoted errors also
incorporating the 9\% calibration error. 

Figs 9 and 10 show the beam-sized  Gaussian convolved survey
maps at 850~${\rm \mu m}$ for ELAIS N2 and the Lockman
Hole East respectively, with the sources $>3.50\sigma$ circled and
numbered. We have also applied our maximum-likelihood estimator to the 
450~${\rm \mu m}$ data (Fox et al. 2002), 
centring normalised 450~${\rm \mu m}$ 
beam maps at the same
positions as in the 850~${\rm \mu m}$ images.  Column 7 gives the 450~${\rm \mu
m}$ flux densities if there is a detection in the short-wavelength data
with S/N $>3.00\sigma$, or, more commonly, a $3\sigma$-upper-limit. 
We also carried out a search for peaks in the short-wavelength
data lying within a search radius of 8 arcsec from the 850~${\rm \mu m}$
source peaks (the combined result of adding in quadrature the 1/2 
beam-width positional uncertainties of the long- and short-wavelength
data) and again employed our source-extraction algorithm. The flux
densities of any detections with S/N better than $3.00 \sigma$ are
given in column 9, with their significances and positional offset
from the long-wavelength position in columns 10 and 11. The errors
quoted on the 450~${\rm \mu m}$ flux densities also include the 20\%
calibration error. 

Given that the FWHM of the JCMT beam at 850~${\rm \mu m}$ is 14.5 arcsec,
the source positions are only likely to be accurate to within
$\simeq 3 - 4$ arcsec. Consistent with this, our simulations suggest 
that an 8~mJy
source, detected at $>3.50\sigma$, should have accurate astrometry to
$\sim 3.5$ arcsec in uniform noise regions and $\sim 4$ arcsec in the
non-uniform noise areas, the positional uncertainty decreasing with
increasing flux. However, these simulations do not account for the
elevation drive pointing glitch mentioned in Section 3.2, and so it is
possible that the astrometric accuracy of a source affected by this
problem could be somewhat worse than this.

Analysis of our fully-simulated survey areas suggests that of the 19
sources with S/N $>4.00$, only one is likely to be spurious. At
the $3.50\sigma$ threshold we have 38 sources, of which $\simeq 7$ 
may be spurious. The situation is markedly worse at $3.00\sigma$, where
$\simeq 30$ of the supposedly-significant peaks may in fact be due to noise or 
confusion of low flux-density sources. 

In the short-wavelength data, we can find only 3 detections at
$>4.00\sigma$ within an 8 arcsec radius of the 850~${\rm  \mu m}$
positions. These 450~${\rm \mu m}$ detections appear to be associated with
the most significant Lockman Hole East 850~${\rm \mu m}$ source and the two
most significant ELAIS N2 850~${\rm  \mu m}$ sources. There are additionally
another five 450~${\rm \mu m}$ detections at better than $3.00\sigma$, of which
all but one is associated with a $>3.50\sigma$ 
850~${\rm  \mu m}$ source. We
must be cautious when considering the credibility of the lower significance
($>3.00\sigma$) 450~${\rm \mu m}$ detections. The atmospheric opacity at 
450~${\rm \mu m}$ is much more 
sensitive to changes in the amount of water vapour
present than at 850~${\rm  \mu m}$, and so the resulting short-wavelength
maps are noisier than their long-wavelength counterparts. Furthermore,
the smaller beam-size (7.5 arcsec FWHM at 450~${\rm  \mu m}$ as opposed to
14.5 arcsec FWHM at 850~${\rm  \mu m}$) means that the short-wavelength
maps will contain a greater number of spurious detections. For S/N
$>3.00$ we would expect 53 spurious sources across the full 260
$\mathrm{arcmin^{2}}$  of short-wavelength data ( and 14 spurious
detections in the long-wavelength data) based on Gaussian statistics alone.

A detailed discussion of individual sources may be found in paper II
of the SCUBA 8~mJy survey (Fox et al., 2002), along with a
multiwavelength analysis considering possible identifications and
SED-based redshift constraints. The 450~${\rm  \mu m}$ images of the survey
fields are also presented by Fox et al. (2002).

\begin{table*}
\begin{tabular}{|c|c|c|c|c|c|} \hline
Flux density & Number & N($>\mathrm{S}$) & Number after & Number after
& N($>\mathrm{S}$)  \\
/mJy & observed & observed & flux boost & flux boost \& & corrected \\
 & &  & correction & completeness &  \\
 & & & & correction & \\ \hline 
\phantom{1}5.0 & 3 & 31 & 0 & 0.00 & 32.9 \\
\phantom{1}5.5 & 4 & 28 & 1 & 3.59 & 32.9 \\
\phantom{1}6.0 & 2 & 24 & 0 & 0.00 & 29.3 \\
\phantom{1}6.5 & 1 & 22 & 3 & 6.55 & 29.3 \\
\phantom{1}7.0 & 1 & 21 & 3 & 5.66 & 22.7 \\
\phantom{1}7.5 & 1 & 20 & 0 & 0.00 & 17.1 \\
\phantom{1}8.0 & 2 & 19 & 3 & 4.64 & 17.1 \\
\phantom{1}8.5 & 4 & 17 & 1 & 1.44 & 12.4 \\
\phantom{1}9.0 & 3 & 13 & 1 & 1.36 & 11.0 \\
\phantom{1}9.5 & 1 & \phantom{1}9 & 0 & 0.00 &   \phantom{1}9.6 \\
10.0 & 1 & \phantom{1}9 & 3 & 3.75 &  \phantom{1}9.6 \\
10.5 & 2 & \phantom{1}8 & 2 & 2.42 &  \phantom{1}5.8 \\
11.0 & 3 & \phantom{1}6 & 0 & 0.00 &  \phantom{1}3.4 \\
11.5 & 1 & \phantom{1}3 & 1 & 1.15 &  \phantom{1}3.4 \\
12.0 & 1 & \phantom{1}2 & 1 & 1.13 &  \phantom{1}2.2 \\
12.5 & 1 & \phantom{1}1 & 1 & 1.11 &  \phantom{1}1.1 \\
13.0 & 0 & \phantom{1}0 & 0 & 0.00 &  \phantom{1}0.0 \\ \hline
\end{tabular}
\caption{\small A breakdown of the raw and corrected source counts at
each 0.5~mJy interval bin for the sources with S/N $>3.50$ from both
survey maps, excluding
those uncovered in the non-uniform noise regions. Column 1 gives 
the flux density, and
columns 2 and 3 the raw number counts in the 0.5~mJy wide bin and
cumulative to that flux-density level. Column 4 gives the number in each bin
once flux-density boosting has been accounted for using the fitted curve in
Fig 4(a), and similarly column 5 gives the number in each bin
once completeness has also been taken into account using the fitted
curve in Fig 2(a). Column 6 gives the cumulative corrected
counts down to each flux-density level.}
\end{table*}

\begin{table*}
\begin{tabular}{|c|c|c|} \hline
Flux density & Raw 850~${\rm \mu m}$ source counts & Corrected 850~${\rm \mu
m}$ source counts \\
/mJy & $N (>S) \mathrm{deg^{-2}}$ & $N (>S) \mathrm{deg^{-2}}$ \\ \hline
\phantom{1}5 & $580 \pm 100$ & $620^{+110}_{-190}$ \\
\phantom{1}6 & $450 \pm 90$\phantom{0} & $550^{+100}_{-170}$ \\
\phantom{1}7 & $400 \pm 90$\phantom{0} & $430^{+90}_{-120}$ \\
\phantom{1}8 & $360 \pm 80$\phantom{0} & $320^{+80}_{-100}$ \\
\phantom{1}9 & $250 \pm 70$\phantom{0} & $200^{+60}_{-70\phantom{0}}$ \\
10 & $170 \pm 60$\phantom{0} & $180^{+60}_{-60\phantom{0}}$ \\
11 & $110 \pm 50$\phantom{0} & \phantom{1}$60^{+30}_{-30\phantom{0}}$ \\
12 & \phantom{1}$40 \pm 30$\phantom{0} & \phantom{1}$40^{+30}_{-30\phantom{0}}$ \\ \hline
\end{tabular}
\caption{\small The 850~${\rm  \mu m}$ source counts per square
degree based on sources with S/N $>3.50$ in both survey maps, and excluding those detected
in the non-uniform noise regions. Column 1 gives the flux density and column 2 the cumulative raw counts per
square degree with the Poisson error. Column 3 gives the cumulative corrected
counts per square degree, the upper error corresponding to the Poisson
error, and the lower error accounting for both the Poisson error and
the presence of spurious sources based on the simulation data.}
\end{table*}

\section{Number counts}

In order to obtain the best estimate of the number counts at 850~${\rm  \mu
m}$, we must account for the effects of flux-density boosting, incompleteness
and contamination from spurious sources. Using the results of our
simulations carried out on the real survey data, the observed flux
densities of the $>3.50\sigma$ sources were corrected statistically
using the output flux density versus input flux density 
plots shown in Fig 4,
for the appropriate image noise region. The number of boost-corrected
sources in bins 0.5~mJy wide, centred on 5.0, 5.5, 6.0~mJy and so on up to
15~mJy, were then amended for completeness using the best fit
results of the real-data simulations shown in Fig 2. The
integrated source counts were obtained by summing the number of
sources in the bins down to the required threshold flux. Since the
scatter in the simulation results is substantially higher in the
non-uniform noise regions (reflected by reduced $\chi^{2}$ values 
of the plotted best-fit curves being a factor of $\sim 5$ higher in
the edge regions),  we have based our estimation 
of the 850~${\rm \mu m}$ source counts purely on the sources detected 
in the areas of uniform
noise, covering an area of 190 $\mathrm{arcmin^{2}}$. In practice this
means that 5/38 of the sources above the $3.50\sigma$ significance 
threshold have been excluded.
This process is summarised in Table 7, with the raw and amended
counts per square degree shown in Table 8, along with Poisson errors. 

The applied correction
factors have been determined from simulations carried out on the real
images and so are independent of source-count models. We cannot,
however, attempt to correct for the presence of random detections
without making some assumptions about the background
counts. Consequently, the estimated number of spurious sources is
included in the lower error bars combined in quadrature with the
Poisson error, rather than being applied directly
to the actual tabulated source-count values.

Fig 11 shows our corrected source counts, at 1~mJy intervals, from 
5 to 12~mJy, along with data points from other SCUBA surveys (Barger,
Cowie \& Sanders 1999, Blain et al. 1999, Eales et al. 2000, Hughes et al 1998, and
Smail et al. 1997). The upper solid and dashed lines in Fig 11 are 
source-count
models, derived by interpolating the 60~$\rm \mu m$ luminosity function
(Saunders et al., 1990) to 850~$\rm \mu m$, with an assumed dust temperature
of 40K and dust emissivity index $\beta = 1.2$ (consistent with
measurements of local ULIRGs such as Arp220 eg. Dunne et al. 2000), and assuming pure luminosity
evolution of the form $L(z)=L(0)(1+z)^{3}$ to $z=2.0$ (beyond which
the luminosity function is simply frozen). The
solid line assumes an Einstein-de Sitter cosmology, whereas the
dashed line assumes the cosmology $\Omega_{M}=0.3$ and
$\Omega_{\Lambda}=0.7$ (both adopting $H_{0}=67$~${\rm kms^{-1}Mpc^{-1}}$). 
The lower solid and dashed lines
represent the same interpolated 60~$\rm \mu m$ luminosity function with no
luminosity evolution included, again for cosmologies
$\Omega_{M}=1.0$, $\Omega_{\Lambda}=0.0$, and
$\Omega_{M}=0.3$, $\Omega_{\Lambda}=0.7$ respectively.
The dotted line shows the adopted source counts model, which was
assumed when creating the fully-simulated images.
The dot-dash and dot-dot-dash lines represent a more physical form of
luminosity evolution (Jameson 2000, Smail et al. 2001), which is compatible with
models of cosmic chemical evolution and includes a peak in
the evolution function:
\be L(z)=L(0)(1+z)^{3/2}\mathrm{sech^{2}[b~ln(1+\mathit{z}\mathrm) -c] cosh^{2}c} \ee
where $\mathrm{b=2.2 \pm 0.1}$ and $\mathrm{c=1.84 \pm 0.1}$. The
dot-dash line assumes a cosmology $\Omega_{M}=1.0$,
$\Omega_{\Lambda}=0.0$, and the dot-dot-dash line $\Omega_{M}=0.3$,
$\Omega_{\Lambda}=0.7$. A dust emissivity index of $\beta = 1.2$ and a
dust temperature of 37K were also assumed (Smail et al. 2001).

\section{2-point correlation functions}

In order to obtain a first measure of the clustering properties of each
of our two fields, we have used our $>3.50\sigma$ sources to calculate
2-point angular correlation functions for both ELAIS N2 and the
Lockman Hole East. We created a catalogue for each survey area,
containing $\sim 40000$ fake sources placed according to a
Poisson distribution within the boundaries of the real data. Although
the positions were generated at random, we used our uncorrelated noise
maps to weight the number density of sources across the image, since
we would expect to uncover a larger density of sources above a
specified signal-to-noise threshold in areas of lower noise. In
practise, we divided the full image into a series of sub-images, 20
arcsec by 20 arcsec in size, and calculated the mean noise level
in each of these grid sections. For sources brighter than $\simeq 5$~mJy, the number of sources expected above a constant signal-to-noise
threshold will increase approximately as $N(>S) \propto S^{-1.5}$, and
consequently we would expect the relative number of sources in each
sub-image to scale as $\langle \mathrm{noise} \rangle^{-1.5}$.

The angular correlation function $w(\theta)$ describes the excess
probability of finding a source at an angular separation $\theta$ from
a source selected at random, over a random distribution. There are a
variety of possible estimators for $w(\theta)$ as a function of 
pair-count ratios. Here we have used:
\be 1+w(\theta) = \frac{2m}{(n-1)}\frac{\langle DD \rangle}{\langle DR
\rangle} \ee
where there are $n$ and $m$ objects in the real and mock catalogues
respectively, $DD$ is the number of distinct data pairs in the real
image within a bin covering a specified range of $\theta$, and $DR$ is
the number of cross-pairs between the real and fake catalogues within
the same range of $\theta$. 

The angular correlation functions may be seen with Poisson errors in
Figs 12 and 13, where a bin size of twice the beam (29.0 arceconds) has been
used. 

\begin{figure*}
\begin{centering}
\centerline{\epsfig{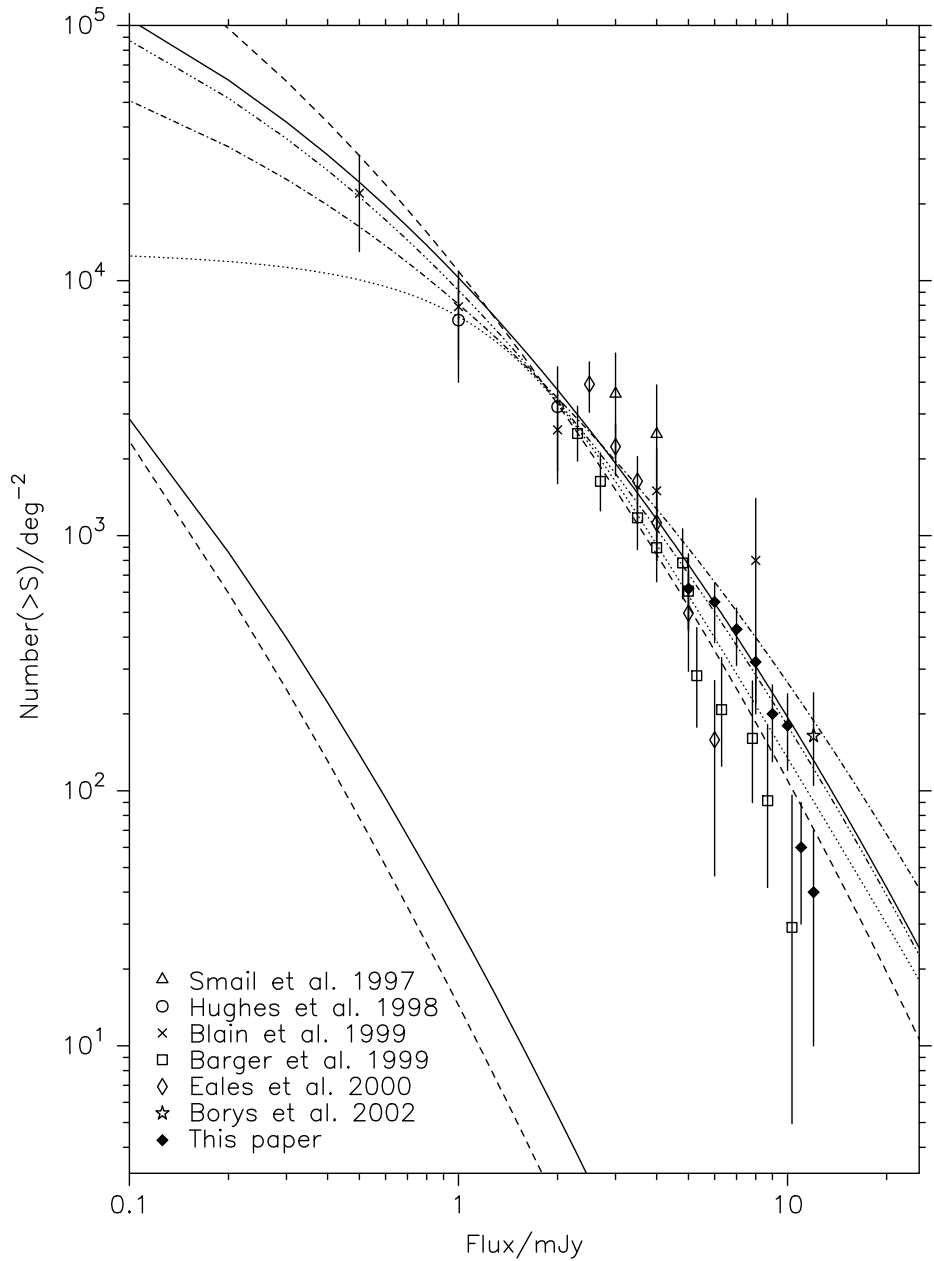}} 
\caption{\small Cumulative 850~$\rm \mu m$ source counts. The solid
diamonds show the new results from the `SCUBA 8~mJy Survey' derived in this paper. The
upper solid and dashed curves are predicted number counts, based
on the 60~$\rm \mu m$ luminosity function of Saunders et al. (1990),
assuming a dust temperature of 40 K, a dust emissivity index of
$\beta=1.2$, and pure luminosity evolution of the form
$L(z)=L(0)(1+z)^{3}$ out to $z=2$ (beyond which the luminosity function
is simply frozen),
for cosmologies ($\Omega_{M}=1.0$,$\Omega_{\Lambda}=0.0$) and
($\Omega_{M}=0.3$,$\Omega_{\Lambda}=0.7$) respectively. The lower
solid and dashed curves show the number counts predicted by the same
models if no luminosity evolution is included. The dotted line shows
the counts model assumed when creating simulated images. The dot-dash and dot-dot
dash assume a more complicated, but arguably more realistic 
luminosity evolution of the form
$L(z)=L(0)(1+z)^{3/2}\mathrm{sech^{2}[b~ln(1+\mathit{z}\mathrm) -c]
cosh^{2}c}$ , for cosmologies
($\Omega_{M}=1.0$,$\Omega_{\Lambda}=0.0$) and
($\Omega_{M}=0.3$,$\Omega_{\Lambda}=0.7$) respectively, and assuming a
dust temperature of 37K and dust emissivity index of $\beta=1.2$
(Jameson 2000, Smail et al. 2002).}
 \end{centering}
\end{figure*}

\section{Discussion}
\subsection{Comparative Source Counts and Models}

Although our survey was undertaken with the aim of constraining the brighter
end of the 850~$\rm \mu m$ number counts, we have in fact 
achieved the most accurate determination to date of the sub-mm source
counts down to $S_{850} \simeq 5$~mJy. From Fig 11 we see that at 5~mJy
there is good agreement between our counts $620^{+110}_{-190}
\mathrm{deg^{-2}}$ and the values of Eales et al. 
(2000) $500 \pm 200 \mathrm{deg^{-2}}$, and
Barger, Cowie \& Sanders (1999) $610^{+240}_{-190} \mathrm{deg^{-2}}$, to well within the quoted
errors. However, at larger flux densities, the number of sources per
square degree in both the Eales and Barger survey areas is somewhat
lower than our counts. For example at 6~mJy the number count of Eales
et al. is a factor of $\sim 3$ lower than our value, and that of
Barger, Cowie \& Sanders (1999) is lower by 
a factor of $\sim 2.5$. The most likely explanation for this
discrepancy is that it is due to small number statistics rather than a
real steepening of the source counts at $> 6$~mJy. We would in fact
\emph{expect this to be the case} if SCUBA sources were generally
clustered on scales of a few arcmin (as indicated by our two-point 
autocorrelation functions, albeit with large errors, 
shown in Section 8 and discussed in Section 9.4). The ``SCUBA 8~mJy Survey'' covers 
260 $\mathrm{arcmin^{2}}$ of sky, 5
times the area of the `Canada UK Deep Sub-millimetre Survey (CUDSS)
14h Field' of Eales et al., and $\sim 2.5$ times the area of Barger
et al., making us less susceptible to arcmin-scale clustering and
thus leaving us much better able to constrain the number counts at
brighter flux densities. At 8~mJy, Blain et al. (1999) suggest a rather 
higher count of $800 \pm 600 \mathrm{deg^{-2}}$  
than our value of $320^{+80}_{-100}
\mathrm{deg^{-2}}$, however the large error on the Blain et
al. number count renders their result almost meaningless. 
At flux densities $> 10$~mJy our survey also inevitably becomes 
limited by the lower surface density of brighter sources. In a similar 
manner to the source counts of Eales et al. and Barger et al. beyond 6~mJy, 
our counts at 11 and 12~mJy suggest a steepening of the counts slope 
towards brighter fluxes. Borys et al. (2002) report a number count of
$164^{+78}_{-59} \mathrm{deg^{-2}}$ at 12~mJy,
suggesting that this is again a statistical rather than a real effect. 

The source counts accumulated across all the sub-millimetre surveys
(as shown in Fig 11) are consistent with a population of high-redshift 
($z>1$), dusty,
star-forming galaxies, analogous to the local ULIRGs, with dust
temperatures $T_{d} \sim 40$K, emissivity $\beta \sim 1.2$, and pure
luminosity evolution of the form $(1+z)^{3}$ out to a redshift
$z=2$ (and simply frozen beyond). We also find that our source counts
are in very good agreement with the more physical form of luminosity
evolution of Jameson (2000) and Smail et al. (2002), 
assuming a dust temperature
of 37 K, particularly for the cosmology
$\Omega_{M}=0.3$, $\Omega_{\Lambda}=0.7$. The lower solid and dashed
lines show the predicted source counts using the same luminosity function but
without any luminosity evolution. Such a model is inconsistent with
the measured number counts by $2-3$ orders of magnitude, which simply
serves to re-emphasise the extent of the evolution in the starburst
population uncovered by SCUBA.

\subsection{Dust Masses and Star Formation Rates}

The photometric redshift estimates of the 19 sources discussed in
Fox et al. (2002) and Lutz et al. (2001) are all indicative of redshifts
$z>1$, and in the vast majority of cases favour $z > 2$. 
Even without knowing a precise redshift, this fact coupled with
an almost uniform sensitivity to objects between redshifts $z=1$ and
$z=8$, allows us to make an estimate of the dust-enshrouded SFRs and
dust masses. Assuming a dust temperature of 40K
and emissivity index $\beta=1.2$, we have interpolated the inferred 850~$\rm \mu m$
luminosity to 60~$\rm \mu m$ using an optically-thin greybody
approximation. The far-infrared luminosity provides a measure of the
current SFR of massive stars. In regions of intense star formation,
dust is heated primarily by the embedded O and B stars which evolve rapidly and
dispense their surrounding material on similarly short time-scales
($\sim 10^{7}$ yr, Wang 1991), hence the rate of dust production is
proportional to the star-formation rate. We may calculate the
SFR to within a factor of a few, by means of
\be \mathrm{SFR} = \epsilon 10^{-10} \mathrm{\frac{L_{60}}{L_{\odot}}
M_{\odot} yr^{-1}}\ee  
The value of $\epsilon$ is uncertain to within a factor
of $\sim 3$ (Scoville \& Young 1983, Thronson \& Telesco 1986,
Rowan-Robinson et al. 1997, Rowan-Robinson 2000), the main sources of
error 
arising from uncertainties in the initial mass function
(IMF), the fraction of optical/UV light absorbed by the dust, and the
time-scale of the burst. We have assumed a value of $\epsilon =2.1$
(Thronson \& Telesco 1986), which incorporates a burst of O,B, and A
type star-formation over $\sim 2 \times 10^{6}$ yr, and assumes a
Salpeter IMF. It was also assumed that all of the optical/UV radiation
was absorbed and re-radiated thermally by the surrounding dust.

\begin{figure*}
\begin{centering}
\centerline{\epsfig{
file=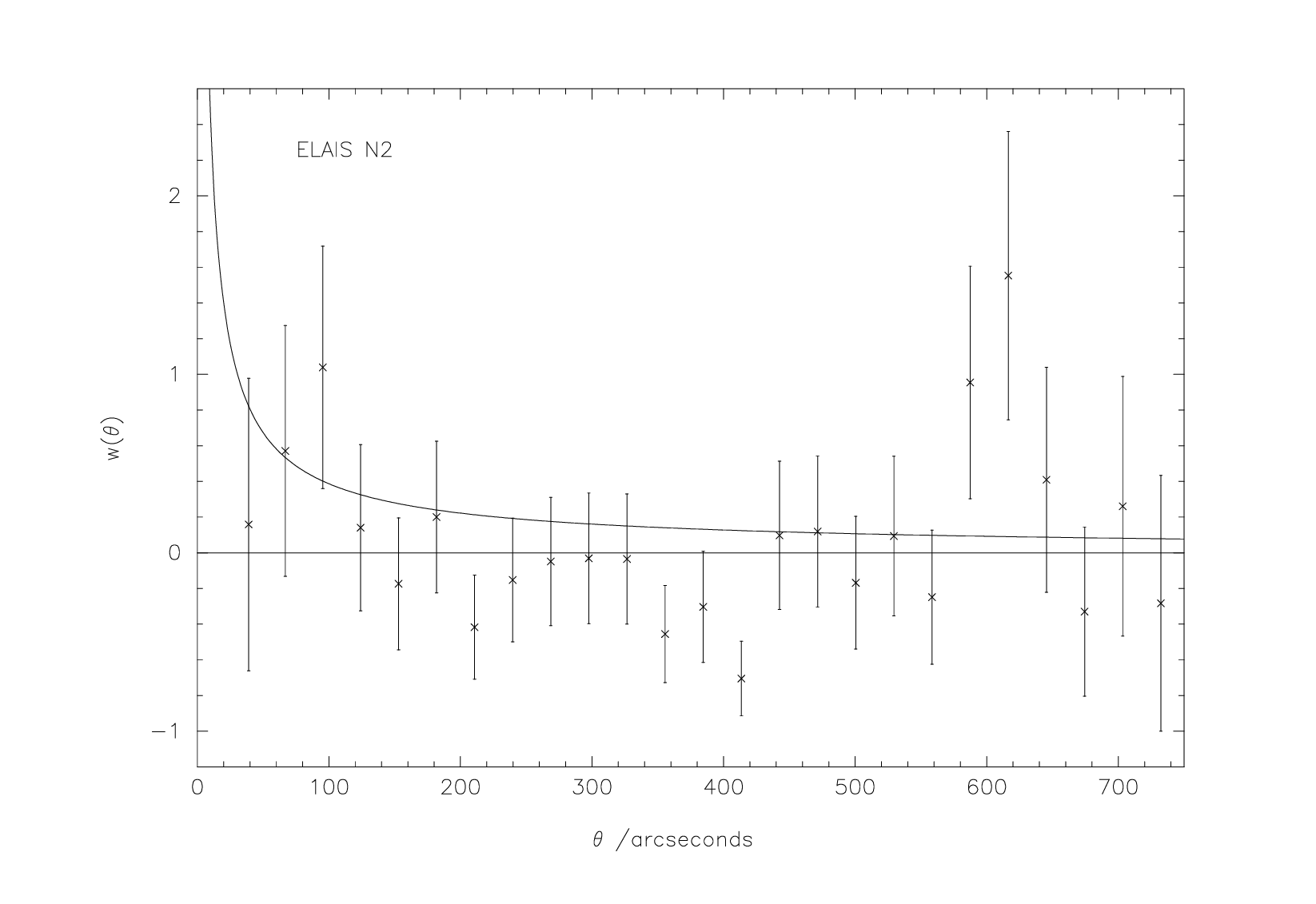,
width=9.6cm,angle=270,clip=}} 
\caption{\small 2-point angular correlation function for
the ELAIS N2 survey field. $w(\theta)=\frac{2m}{n-1}\frac{\langle
DD\rangle}{\langle DR\rangle}
- 1$
where there are n and m real and fake $>3.50\sigma$ sources
respectively, and $DD$ and $DR$ are the number of distinct data pairs
and cross data-random pairs within 29 arcsec width bins as a
function of angular separation $\theta$. The power-law line indicates the 
correlation function found by Daddi et al. (2000) for EROs with $R-K > 5$
and $K < 18.5$ as discussed in Sections 9.3 and 9.4.}
\end{centering}
\end{figure*}

\begin{figure*}
\begin{centering}
\centerline{\epsfig{
file=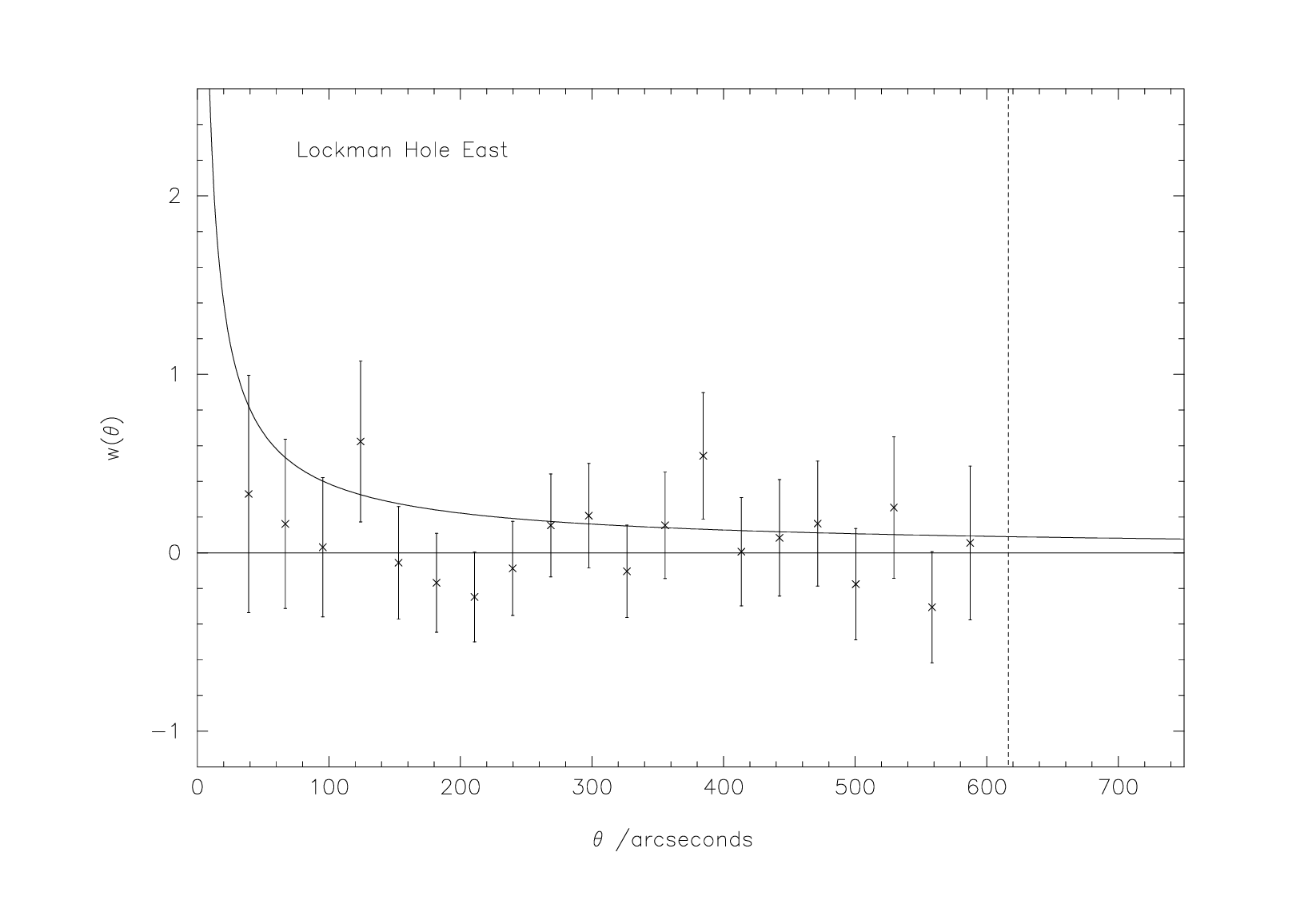,
width=9.6cm,angle=270,clip=}} 
\caption{\small 2-point angular correlation function for
the Lockman Hole East survey
field. $w(\theta)=\frac{2m}{n-1}\frac{\langle DD\rangle}{\langle DR\rangle} - 1$
where there are n and m real and fake $>3.50\sigma$ sources
respectively, and $DD$ and $DR$ are the number of distinct data pairs
and cross data-random pairs within 29 arcsec width bins as a
function of angular separation $\theta$.
The power-law line indicates the 
correlation function found by Daddi et al. (2000) for EROs with $R-K > 5$
and $K < 18.5$ as discussed in Sections 9.3 and 9.4. Note that the
vertical dashed line demonstrates the cut-off at which there are no
more source pairings in the real data.}
\end{centering}
\end{figure*}

In calculating the mass of dust present in each source, we assumed the
sub-millimetre continuum to be the result of optically-thin thermal
emission from the heated dust grains, with no additional contribution
from bremsstrahlung or synchrotron radiation. We may then determine
the dust mass $M_{d}$ by means of the relation
\be M_{d} = \frac{1}{1+z}
\frac{S_{obs}D_{L}^{2}}{k_{d}^{\mathrm{rest}}B(\nu^{\mathrm{rest}},T_{d})}
\ee
where $z$ is the redshift of the source, $S_{obs}$ is the observed
flux density, $D_{L}$ is the luminosity distance,
$k_{d}^{\mathrm{rest}}$ is the rest-frequency mass absorption
coefficient, and $B(\nu^{\mathrm{rest}},T_{d})$ is the rest-frequency
value of the Planck function from dust grains radiating at temperature
$T_{d}$. The main uncertainty in determining the dust mass $M_{d}$ is
the uncertainty in the rest-frequency mass absorption
coefficient $k_{d}^{\mathrm{rest}}$. We have adopted the same approach
as Hughes, Dunlop \& Rawlings (1997), using their average value of
$k_{d}$(800~$\rm \mu m$)$ = 0.15 \pm 0.09$~${\rm m^{2}kg^{-1}}$, and
interpolating to shorter wavelengths by assuming 
$k_{d}~\propto~\lambda^{-1.5}$. 
A different choice of $k_{d}$(800~$\rm \mu m$) would be
expected to change the dust mass estimates by a factor of up to 2.

The results are outlined in Tables 9 and 10, the
un-bracketted quantities assuming an Einstein-de Sitter cosmology, and
the values enclosed in brackets adopting 
$\Omega_{M}=0.3$, $\Omega_{\Lambda}=0.7$.
SFRs range from  several hundred to several thousand solar masses
per year (exceeding even that of the extreme local starburst Arp220),
the star-forming activity being heavily obscured by $10^{8}-10^{9}~{\rm 
M_{\odot}}$ of dust. In fact, for $\Omega_{M}=0.3$,
$\Omega_{\Lambda}=0.7$ and $H_0 = 67$~${\rm km s^{-1} Mpc^{-1}}$
it can be seen that an 850~$\rm \mu m$ flux density 
of 8~mJy corresponds closely to a star-formation rate of $1000~{\rm M_{\odot} yr^{-1}}$.

The 19 sources brighter than 8~mJy, with $S/N>3.5$ and
located in the uniform noise regions of the survey maps, account for $\simeq 10\%$ of the
sub-millimetre background observed by COBE-FIRAS 
(Puget et al. 1996, Fixsen et al. 1998, Hauser et al. 1998). Table 11
shows the inferred star formation rate density for a variety of
redshift bands, and for both of our adopted cosmologies. These bright
SCUBA sources alone imply a high-redshift star formation rate density (SFRD)
in the range $\simeq 0.01-0.07$ $\mathrm{M_{\odot}yr^{-1}Mpc^{-3}}$ , comparable
to that observed in the optical/UV (Madau et al. 1996, Steidel et al. 1999). If we assume that the whole of the
sub-millimtere background may be attributed to starlight reprocessed
by dust, then this would imply a high-redshift SFRD in the range
$\simeq 0.1-0.7$ $\mathrm{M_{\odot}yr^{-1}Mpc^{-3}}$, varying only by a factor
of $2-3$ on the adopted redshift band for a given choice of cosmology.
Our results agree very well with Barger, Cowie \& Richards (2000) who
considered the contribution to the SFRD of sources brighter than
6~mJy, and determined completeness corrected SFRDs
$\simeq 0.1-0.4$ $\mathrm{M_{\odot}yr^{-1}Mpc^{-3}}$
in the redshift range $1<z<3$, and $\simeq 0.1-0.7$
$\mathrm{M_{\odot}yr^{-1}Mpc^{-3}}$ in the redshift range $3<z<6$. In contrast to
the marked decline in the SFRD at $z>1$ originally implied by
optical/UV observations (Madau et
al. 1996), sub-millimetre surveys suggest that the SFRD
is either steady or gently increasing to perhaps as far back as
$z=5$. Current redshift constraints (Dunlop 2001a and references
within) suggest that only $10-15\%$ of luminous ($> 4$~mJy)
sub-millimetre sources lie at $z<2$, and that the median redshift of
this population is $z_{\mathrm{med}}\simeq 3$. A peak in the SFRD
around this epoch would not be unexpected given the strong correlation
between black-hole and spheroid mass found at low redshift (Kormendy
\& Gebhardt 2001) and the peak in optical emission from powerful
quasars at $z\simeq 2.5$. However, improved redshift constraints are required
to establish when (and indeed if) the comoving SFRD reached a
maximum.
 
\begin{table*}
\begin{tabular}{|c|c|c|r|c|c|c|r|c|} \hline
Source & RA & DEC & $S_{850}$\phantom{00} & S/N       & Noise Region & $\mathrm{log_{10}L_{60}}$    & SFR\phantom{00}        & $\mathrm{log_{10}M_{dust}}$ \\
        &(J2000) &(J2000) &    /mJy\phantom{00}    & 850~$\rm \mu m$& 850~$\rm \mu m$   &
/$\mathrm{L_{\odot}}$ & /$\mathrm{M_{\odot}yr^{-1}}$ &
        /$\mathrm{M_{\odot}}$ \\ \hline

01 & 16:37:04.3 & 41:05:30 &  $11.2 \pm 1.6$ &  8.59 & uniform & 12.49
(12.83)
& 649 (1430) & 8.62 (8.96) \\
02 & 16:36:58.7 & 41:05:24 &  $10.7 \pm 2.0$ &  6.27 & uniform & 12.47 (12.82)
& 623 (1372)& 8.60 (8.95) \\
03 & 16:36:58.2 & 41:04:42 &  $ 8.5 \pm 1.6$ &  5.86 & uniform & 12.37 (12.71)
& 494 (1087)& 8.50 (8.84)\\
04 & 16:36:50.1 & 40:57:33 &  $ 8.2 \pm 1.7$ &  5.18 & uniform & 12.36 (12.70)
& 480 (1056)& 8.49 (8.83)\\
05 & 16:36:35.6 & 40:55:58 &  $ 8.5 \pm 2.2$ &  4.16 & uniform & 12.37 (12.71)
& 493 (1085)& 8.50 (8.84)\\
06 & 16:37:04.2 & 40:55:45 &  $ 9.2 \pm 2.4$ &  4.13 & uniform & 12.41 (12.75)
& 537 (1182) & 8.54 (8.88)\\
07 & 16:36:39.4 & 40:56:38 &  $ 9.0 \pm 2.4$ &  4.07 & uniform & 12.40 (12.74)
& 523 (1152)& 8.53 (8.87)\\ \hline
08 & 16:36:58.8 & 40:57:33 &  $ 5.1 \pm 1.4$ &  3.82 & uniform & 12.15 (12.49)
& 298 (\phantom{1}656) & 8.28 (8.62)\\
09 & 16:36:22.4 & 40:57:05 &  $ 9.0 \pm 2.5$ &  3.76 & non-uniform &
12.40 (12.74)& 526 (1159)& 8.53 (8.87)\\
10 & 16:36:48.8 & 40:55:54 &  $ 5.4 \pm 1.5$ &  3.69 & uniform & 12.18 (12.52)
& 315 (\phantom{1}694)& 8.31 (8.65)\\
11 & 16:36:44.5 & 40:58:38 &  $ 7.1 \pm 2.0$ &  3.67 & uniform & 12.29 (12.63)
& 411 (\phantom{1}904)& 8.42 (8.76)\\
12 & 16:37:02.5 & 41:01:23 &  $ 5.5 \pm 1.6$ &  3.65 & uniform & 12.18 (12.52)
& 319 (\phantom{1}702)& 8.31 (8.65)\\
13 & 16:36:31.2 & 40:55:47 &  $ 6.3 \pm 1.9$ &  3.56 & uniform & 12.24 (12.58)
& 365 (\phantom{1}804)& 8.37 (8.71)\\
14 & 16:36:19.7 & 40:56:23 &  $11.2 \pm 3.3$ &  3.55 & non-uniform &
12.49 (12.83)& 651 (1434)& 8.62 (8.96)\\
15 & 16:37:10.2 & 41:00:17 &  $ 5.0 \pm 1.5$ &  3.52 & uniform & 12.14 (12.48)
& 290 (\phantom{1}639)& 8.27 (8.61)\\
16 & 16:36:52.3 & 41:05:52 &  $12.9 \pm 3.9$ &  3.51 & non-uniform &
12.55 (12.90)& 751 (1655)& 8.68 (9.03)\\
17 & 16:36:51.4 & 41:05:06 &  $ 5.7 \pm 1.7$ &  3.50 & uniform & 12.20 (12.54)
& 331 (\phantom{1}728)& 8.33 (8.67)\\ \hline \hline
18 & 16:36:11.4 & 40:59:26 &  $20.8 \pm 6.2$ &  3.49 & non-uniform &
12.76 (13.10)& 1212 (2669)& 8.89 (9.23)\\
19 & 16:36:35.9 & 41:01:38 &  $ 9.4 \pm 2.8$ &  3.49 & uniform & 12.42 (12.76)
& 550 (1211)& 8.55 (8.89)\\
20 & 16:36:49.4 & 41:04:17 &  $ 6.6 \pm 2.0$ &  3.48 & uniform & 12.26 (12.60)
& 383 (\phantom{1}844)& 8.39 (8.73)\\
21 & 16:37:10.5 & 41:00:49 &  $ 4.9 \pm 1.5$ &  3.47 & uniform & 12.13 (12.47)
& 283 (\phantom{1}624)& 8.26 (8.60)\\
22 & 16:36:27.9 & 40:54:04 &  $13.4 \pm 4.1$ &  3.46 & non-uniform &
12.57 (12.91)& 780 (1718)& 8.70 (9.04)\\
23 & 16:37:19.5 & 41:01:38 &  $12.4 \pm 3.8$ &  3.45 & non-uniform &
12.54 (12.88)& 721 (1587)& 8.67 (9.01)\\
24 & 16:36:26.9 & 41:02:23 &  $10.4 \pm 3.2$ &  3.43 & non-uniform &
12.46 (12.80)& 607 (1336)& 8.59 (8.93)\\
25 & 16:36:18.3 & 40:59:12 &  $12.1 \pm 3.8$ &  3.37 & non-uniform &
12.52 (12.87)& 703 (1548)& 8.65 (9.00)\\
26 & 16:36:32.0 & 41:00:05 &  $ 9.7 \pm 3.1$ &  3.26 & uniform & 12.43 (12.77)
& 567 (1248)& 8.56 (8.90)\\
27 & 16:36:48.3 & 41:03:52 &  $ 6.6 \pm 2.1$ &  3.25 & uniform & 12.26 (12.61)
& 386 (\phantom{1}851)& 8.39 (8.74)\\
28 & 16:36:47.2 & 41:04:48 &  $ 6.3 \pm 2.0$ &  3.24 & uniform & 12.24 (12.59)
& 367 (\phantom{1}809)& 8.37 (8.72)\\
29 & 16:36:24.1 & 40:59:35 &  $ 9.9 \pm 3.3$ &  3.14 & uniform & 12.44 (12.78)
& 573 (1263)& 8.57 (8.91)\\
30 & 16:37:07.5 & 41:02:37 &  $ 4.6 \pm 1.5$ &  3.13 & uniform & 12.11 (12.45)
& 268 (\phantom{1}590)& 8.24 (8.58)\\
31 & 16:36:28.1 & 41:01:41 &  $ 6.8 \pm 2.3$ &  3.07 & uniform & 12.27 (12.62)
& 393 (\phantom{1}866)& 8.40 (8.75)\\
32 & 16:36:39.8 & 41:00:34 &  $ 5.4 \pm 1.8$ &  3.07 & uniform & 12.18 (12.52)
& 315 (\phantom{1}694)& 8.31 (8.65)\\
33 & 16:36:50.5 & 40:58:54 &  $ 4.8 \pm 1.6$ &  3.06 & uniform & 12.12 (12.47)
& 280 (\phantom{1}616)& 8.25 (8.60)\\
34 & 16:36:27.0 & 40:58:15 &  $ 6.9 \pm 2.3$ &  3.05 & uniform & 12.28 (12.62)
& 400 (\phantom{1}881)& 8.41 (8.75)\\
35 & 16:36:44.8 & 40:56:51 &  $ 5.5 \pm 1.9$ &  3.02 & uniform & 12.18 (12.52)
& 318 (\phantom{1}700)& 8.31 (8.65)\\
36 & 16:36:52.9 & 41:02:52 &  $ 3.8 \pm 1.3$ &  3.00 & uniform & 12.02 (12.37)
& 222 (\phantom{1}489)& 8.15 (8.50)\\ \hline
\end{tabular}
\label{table:n2_sfr_dust}\caption{\small Inferred FIR/60~$\rm \mu m$
luminosities, star-formation rates and dust masses for the sources in
ELAIS N2. The un-bracketted numbers in columns 7, 8 and 9 assume an
Einstein - de Sitter cosmology, whereas the numbers enclosed in brackets
assume $\Omega_{M}=0.3$ and $\Omega_{\Lambda}=0.7$. A dust
temperature of 40K, and emissivity index $\beta=1.2$ were assumed
throughout, as was $H_{0}=67$~${\rm kms^{-1}Mpc^{-1}}$.}
\end{table*}

\begin{table*}
\begin{tabular}{|c|c|c|r|c|c|c|r|c|} \hline
Source & RA & DEC & $S_{850}$\phantom{00} & S/N       & Noise Region & $\mathrm{log_{10}L_{60}}$    & SFR\phantom{00}        & $\mathrm{log_{10}M_{dust}}$ \\
        &(J2000) &(J2000) &    /mJy\phantom{00}    & 850~$\rm \mu m$& 850~$\rm \mu m$   &
/$\mathrm{L_{\odot}}$ & /$\mathrm{M_{\odot}yr^{-1}}$ &
        /$\mathrm{M_{\odot}}$ \\ \hline

01 & 10:52:01.4 & 57:24:43 &  $10.5 \pm 1.6$ &  8.10 & deep strip &
12.47 (12.81)& 613 (1350)& 8.60 (8.94)\\
02 & 10:52:38.2 & 57:24:36 &  $10.9 \pm 2.4$ &  5.22 & uniform &
12.48 (12.82)& 635 (1399)& 8.61 (8.95)\\
03 & 10:51:58.3 & 57:18:01 &  $ 7.7 \pm 1.7$ &  5.06 & deep strip &
12.33 (12.67)& 448 (\phantom{1}987)& 8.46 (8.80)\\
04 & 10:52:04.1 & 57:25:28 &  $ 8.3 \pm 1.8$ &  5.03 & deep strip &
12.36 (12.71)& 484 (1067)& 8.49 (8.84)\\
05 & 10:51:59.3 & 57:17:18 &  $ 8.6 \pm 2.0$ &  4.57 & deep strip &
12.38 (12.72)& 500 (1100)& 8.51 (8.85)\\
06 & 10:52:30.6 & 57:22:12 &  $11.0 \pm 2.6$ &  4.50 & uniform & 12.48 (12.83)
& 638 (1405)& 8.61 (8.96)\\
07 & 10:51:51.5 & 57:26:35 &  $ 8.1 \pm 1.9$ &  4.50 & deep strip &
12.35 (12.70)& 474 (1044)& 8.48 (8.83)\\
08 & 10:52:00.0 & 57:24:21 &  $ 5.1 \pm 1.3$ &  4.38 & deep strip &
12.15 (12.49)& 295 (\phantom{1}649)& 8.28 (8.62)\\
09 & 10:52:22.7 & 57:19:32 &  $12.6 \pm 3.2$ &  4.20 & uniform & 12.54 (12.88)
& 736 (1620)& 8.67 (9.02)\\
10 & 10:51:42.4 & 57:24:45 &  $12.2 \pm 3.1$ &  4.18 & uniform & 12.53 (12.87)
& 712 (1567)& 8.66 (9.00)\\
11 & 10:51:30.6 & 57:20:38 &  $13.5 \pm 3.5$ &  4.09 & non-uniform &
12.57 (12.92)& 786 (1731)& 8.70 (9.05)\\
12 & 10:52:07.7 & 57:19:07 &  $ 6.2 \pm 1.6$ &  4.01 & deep strip &
12.23 (12.58)& 358 (\phantom{1}789)& 8.36 (8.71)\\ \hline
13 & 10:51:33.6 & 57:26:41 &  $ 9.8 \pm 2.8$ &  3.69 & uniform & 12.43 (12.78)
& 569 (1252)& 8.56 (8.91)\\
14 & 10:52:04.3 & 57:26:59 &  $ 9.5 \pm 2.8$ &  3.61 & uniform & 12.42 (12.76)
& 552 (1215)& 8.55 (8.89)\\
15 & 10:52:24.6 & 57:21:19 &  $11.7 \pm 3.4$ &  3.60 & uniform & 12.51 (12.85)
& 680 (1497)& 8.64 (8.98)\\
16 & 10:52:27.1 & 57:25:16 &  $ 6.1 \pm 1.8$ &  3.56 & uniform & 12.23 (12.57)
& 356 (\phantom{1}784)& 8.36 (8.70)\\
17 & 10:52:16.8 & 57:19:23 &  $ 9.2 \pm 2.7$ &  3.55 & uniform & 12.41 (12.75)
& 537 (1182)& 8.54 (8.88)\\
18 & 10:51:55.7 & 57:23:12 &  $ 4.5 \pm 1.3$ &  3.55 & deep strip &
12.09 (12.44)& 261 (\phantom{1}575)& 8.22 (8.57)\\
19 & 10:52:29.7 & 57:26:19 &  $ 5.5 \pm 1.6$ &  3.54 & uniform & 12.18 (12.53)
& 321 (\phantom{1}708)& 8.32 (8.66)\\
20 & 10:52:37.7 & 57:20:30 &  $10.3 \pm 3.1$ &  3.51 & non-uniform &
12.46 (12.80)& 599 (1319)& 8.59 (8.93)\\
21 & 10:52:01.7 & 57:19:16 &  $ 4.5 \pm 1.3$ &  3.50 & deep strip &
12.10 (12.44) & 262 (\phantom{1}577)& 8.22 (8.57)\\ \hline \hline
22 & 10:52:05.7 & 57:20:53 &  $ 4.7 \pm 1.4$ &  3.49 & deep strip &
12.11 (12.46)& 272 (\phantom{1}599)& 8.24 (8.59)\\
23 & 10:51:47.0 & 57:24:51 &  $ 7.4 \pm 2.2$ &  3.48 & uniform & 12.31 (12.66)
& 431 (\phantom{1}950)& 8.44 (8.79)\\
24 & 10:51:42.9 & 57:24:12 &  $11.4 \pm 3.4$ &  3.47 & uniform & 12.50 (12.84)
& 661 (1456)& 8.63 (8.97)\\
25 & 10:52:36.0 & 57:18:20 &  $12.7 \pm 3.8$ &  3.46 & non-uniform &
12.55 (12.89)& 740 (1630)& 8.68 (9.02)\\
26 & 10:52:27.3 & 57:19:06 &  $ 8.3 \pm 2.6$ &  3.39 & uniform & 12.36 (12.71)
& 486 (1070)& 8.49 (8.84)\\
27 & 10:51:53.8 & 57:18:47 &  $ 5.5 \pm 1.7$ &  3.38 & deep strip &
12.18 (12.52)& 318 (\phantom{1}701)& 8.31 (8.65)\\
28 & 10:52:34.6 & 57:20:02 &  $10.2 \pm 3.2$ &  3.31 & uniform & 12.45 (12.79)
& 594 (1308)& 8.58 (8.92)\\
29 & 10:52:16.4 & 57:25:07 &  $ 6.7 \pm 2.1$ &  3.30 & uniform & 12.27 (12.61)
& 389 (\phantom{1}856) & 8.40 (8.74)\\
30 & 10:52:42.2 & 57:18:28 &  $11.2 \pm 3.6$ &  3.25 & non-uniform &
12.49 (12.84)& 653 (1437)& 8.62 (8.97)\\
31 & 10:52:03.9 & 57:20:07 &  $ 4.0 \pm 1.3$ &  3.24 & deep strip &
12.05 (12.39)& 234 (\phantom{1}516)& 8.18 (8.52)\\
32 & 10:52:00.0 & 57:20:39 &  $ 4.3 \pm 1.4$ &  3.22 & deep strip &
12.08 (12.42)& 253 (\phantom{1}556)& 8.21 (8.55)\\
33 & 10:51:33.8 & 57:19:29 &  $ 8.1 \pm 2.6$ &  3.20 & uniform & 12.35 (12.69)
& 472 (1040)& 8.48 (8.82)\\
34 & 10:52:09.9 & 57:20:40 &  $ 8.5 \pm 2.8$ &  3.16 & uniform & 12.37 (12.72)
& 495 (1091)& 8.50 (8.85)\\
35 & 10:51:57.6 & 57:26:03 &  $ 6.7 \pm 2.3$ &  3.02 & uniform & 12.27 (12.61)
& 392 (\phantom{1}864)& 8.40 (8.74)\\
36 & 10:52:03.5 & 57:16:54 &  $ 8.0 \pm 2.8$ &  3.00 & uniform & 12.34 (12.69)
& 463 (1019)& 8.47 (8.82)\\ \hline
\end{tabular}
\label{table:lh_sfr_dust}\caption{\small Inferred FIR/60~$\rm \mu m$
luminosities, star-formation rates and dust masses for the sources in
the Lockman Hole East. The un-bracketted numbers in columns 7, 8 and 9
assume an Einstein - de Sitter cosmology, whereas the numbers enclosed
in brackets assume $\Omega_{M}=0.3$ and $\Omega_{\Lambda}=0.7$. 
A dust temperature of 40K,
and emissivity index $\beta=1.2$ were assumed throughout, as was
$H_{0}=67$~${\rm kms^{-1}Mpc^{-1}}$.} 
\end{table*}

\subsection{Co-moving number density of 8~mJy sources}

As explained in the previous Section, although we currently lack 
redshifts for the sub-mm sources uncovered in this survey, existing 
constraints mean that we can be very confident that virtually all of them
lie at $z > 1$, and indeed can be reasonably confident that the majority 
lie at $z > 2$ (Dunlop 2001a, Fox et al. 2002). It is thus still possible to 
make a meaningful 
estimate of the co-moving number density of bright ($S_{850} > 8$~mJy) sub-mm 
sources at high redshift implied by the number counts derived from this
survey. The results of this calculation, for a variety 
of assumed redshift bands,
are summarized in Table 12 (again for our two standard assumed alternative 
cosmologies). From this table it can be seen that the co-moving number
density of dust-enshrouded starburst galaxies with star formation rates $>
1000~{\rm M_{\odot} yr^{-1}}$ lies in the range $\simeq 1 -10 
\times 10^{-5}~{\rm Mpc^{-3}}$ in very good agreement with Lilly et
al. (1999), Barger, Cowie \& Sanders (1999), and Barger, Cowie \& Richards
(2000). 
In a similar manner to the comoving SFRD, this result 
depends only weakly (a factor of $2 - 3$) on the precise choice of 
assumed redshift band for the sources, for a given choice of cosmology. Adopting the now strongly-favoured
flat, $\Lambda$-dominated
cosmology leads us to conclude that the co-moving number density of 
dust-enshrouded starburst galaxies with star formation rates $>
1000~{\rm M_{\odot} yr^{-1}}$ is $\simeq 1 \times 10^{-5}~{\rm Mpc^{-3}}$.

This is an interesting number. It is over an order of magnitude smaller than 
the number density of galaxies brighter than $L^{\star}$ in the present-day 
universe, as inferred from the $K$-band luminosity function (Glazebrook
et al. 1995; Cowie et al. 1995; Gardner et al. 1997; Szokoly et al. 1998; 
Kochanek et al. 2001), but is an order of magnitude greater than 
the co-moving number density of bright optical QSOs ($M_V < -24$) at 
$z \simeq 2 - 3$ (Warren, Hewett \& Osmer 1995).

With reference to the present-day $K$-band luminosity function, a co-moving
number density of $1 \times 10^{-5}~{\rm Mpc^{-3}}$ corresponds to a galaxy
luminosity $1 - 1.5$ magnitudes brighter than $L^{\star}$, or equivalently to 
galaxies $3-4$ times more massive than an evolved $L^{\star}$ galaxy. In this 
regime the present-day galaxy population is completely dominated by massive 
ellipticals (Kochanek et al. 2001). Interestingly the hosts of 
present-day FRII radio
sources, and the more luminous radio-quiet quasars are also confined to this 
same high-mass regime (Dunlop 2001b, Dunlop et al. 2002). 

Thus, if one wants to attempt to link the high-redshift population of very 
luminous (i.e. SFR $ \simeq 1000~{\rm M_{\odot} yr^{-1}}$) dust-enshrouded
starburst galaxies to a low-redshift population, purely on the basis of 
number-density coincidence, then the simplest connection is 
that the bright SCUBA 
galaxies are the progenitors of present-day massive ellipticals with 
stellar masses $\simeq 10^{12} {\rm M_{\odot}}$. This is not
unreasonable, given that such objects require star-formation to be sustained 
at $ \simeq 1000~{\rm M_{\odot} yr^{-1}}$ 
for $\simeq 1$~Gyr to assemble their present-day 
stellar populations. It is also consistent with the discovery that bright 
SCUBA detections of (comparably massive) radio galaxies are largely confined 
to $z > 2$ (Archibald et al. 2001).

If the luminous sub-mm SCUBA sources uncovered by this survey are indeed 
the progenitors of the present-day massive elliptical population
(Dunlop 2001c), and if, as 
current evidence suggests, the majority of bright sub-mm sources lie at 
$z > 2$ (Dunlop 2001a), then one might reasonably expect to find a 
comparably-numerous population of passively-evolving massive ellipticals at 
intermediate
redshifts. One way to search for such a population is via surveys designed
to detect extremely red objects (EROs), such as that recently undertaken by 
Daddi et al. (2000, 2002). One can obtain a rough estimate of the co-moving 
number density of passively-evolving massive ellipticals at intermediate 
redshifts by considering the surface density of EROs in the Daddi et al. 
survey with $R-K > 5.3$ (setting a lower redshift boundary of $z \simeq 1$) 
and $K < 18.5$ (setting an approximate upper redshift boundary of $z \simeq
2$ for ellipticals of comparable mass to bright radio galaxies - Jarvis et al. (2001)). The surface density of such objects is $\simeq 0.1$ per sq. 
arcmin ($\equiv 350$ per sq. degree), very similar to the surface 
density found here
for 8~mJy SCUBA sources. For the appropriate 
redshift band $1 < z < 2$ this surface 
density converts into a co-moving number density of 
$3 \times 10^{-5}~{\rm Mpc^{-3}}$ assuming 
$\Omega_M = 0.3$, $\Omega_{\Lambda} = 0.7$ which, as can 
be seen from Table 12, agrees with the co-moving number density of bright SCUBA sources
to within a factor of 2 or 3 (depending on the choice of
assumed redshift band for the SCUBA sources). 

This numerical coincidence provides further circumstantial
evidence for an evolutionary path for {\it all} massive ellipticals which 
mirrors that which has been already largely established for massive radio 
galaxies (Archibald et al. 2000a; Willott et al. 2001; Jimenez et al. 1999), 
i.e.

\vspace*{0.3cm}

\noindent
SCUBA source at $z \ge 2.5$ $\rightarrow$\\ 
ERO at $z \simeq 1.5$ $\rightarrow$\\ 
$3-4L^{\star}$ evolved elliptical at  $z = 0$

\vspace*{0.3cm}

A key test of this picture will be to establish
whether or not the bright SCUBA sources 
display comparable or even stronger 
{\it spatial} clustering than the EROs, for which Daddi et 
al. (2000) report $r_0 \simeq 11 \, 
h^{-1}{\rm Mpc}$ (for objects
with $R-K > 5$). As described in the next 
Section, there is tantalyzing evidence of 
angular clustering in the 8-mJy survey, but
it is clear that a substantially larger sub-mm survey
(approaching 1 degree in size) 
will be required to settle this issue.

\begin{table*}
\begin{tabular}{|c|c|c|} \hline
Redshift range & Co-moving SFR density & Co-moving SFR density \\
 & ($\Omega_{M}=1.0$, $\Omega_{\Lambda}=0.0$) &
 ($\Omega_{M}=0.3$, $\Omega_{\Lambda}=0.7$) \\ 
 & /$\mathrm{M_{\odot}yr^{-1}Mpc^{-3}}$ &  /$\mathrm{M_{\odot}yr^{-1}Mpc^{-3}}$ \\ \hline
$1-3$ & 0.028 & 0.018 \\ 
$2-3$ & 0.057 & 0.033 \\ 
$2-4$ & 0.030 & 0.017 \\ 
$2-5$ & 0.022 & 0.012 \\ 
$3-4$ & 0.065 & 0.035 \\ 
$3-5$ & 0.035 & 0.018\\ \hline
\end{tabular}
\caption{\small The co-moving star formation rate density of sources brighter than 8~mJy, 
detected with S/N $>3.50$, using the uniform noise regions of the
two survey areas and assuming that all of these sources lie
within the redshift range given in column 1. Column 2 assumes an
Einstein-de Sitter cosmology, and column 3 assumes
$\Omega_{M}=0.3$, $\Omega_{\Lambda}=0.7$. $H_{0}=67$~${\rm kms^{-1}Mpc^{-1}}$ 
was adopted for both cosmologies.}
\end{table*}

\begin{table*}
\begin{tabular}{|c|c|c|} \hline
Redshift range & Co-moving no. density & Co-moving no. density \\
 & ($\Omega_{M}=1.0$, $\Omega_{\Lambda}=0.0$) &
 ($\Omega_{M}=0.3$, $\Omega_{\Lambda}=0.7$) \\ 
 & /$\mathrm{Mpc^{-3}}$ &  /$\mathrm{Mpc^{-3}}$ \\ \hline
$1-3$ & $\phantom{1}4.94 \times 10^{-5}$ & $1.40 \times 10^{-5}$ \\ 
$2-3$ & $\phantom{1}9.96 \times 10^{-5}$ & $2.63 \times 10^{-5}$ \\ 
$2-4$ & $\phantom{1}6.05 \times 10^{-5}$ & $1.34 \times 10^{-5}$ \\ 
$2-5$ & $\phantom{1}3.74 \times 10^{-5}$ & $0.94 \times 10^{-5}$ \\ 
$3-4$ & $11.22 \times 10^{-5}$ & $2.77 \times 10^{-5}$ \\ 
$3-5$ & $\phantom{1}6.00 \times 10^{-5}$ & $1.45 \times 10^{-5}$ \\ \hline
\end{tabular}
\caption{\small The co-moving number density of sources brighter than 8~mJy, 
detected with S/N $>3.50$, using the uniform noise regions of the
two survey areas and assuming that all of these sources lie
within the redshift range given in column 1. Column 2 assumes an
Einstein-de Sitter cosmology, and column 3 assumes
$\Omega_{M}=0.3$, $\Omega_{\Lambda}=0.7$. $H_{0}=67$~${\rm kms^{-1}Mpc^{-1}}$ 
was adopted for both cosmologies.}
\end{table*}

\subsection{Clustering}

If the bright 850~$\rm \mu m$ sources are
indeed the progenitors of massive elliptical galaxies then they should
be strongly clustered, an inevitable result of gravitational collapse
from Gaussian initial density fluctuations since the rare high-mass
peaks are strongly biased with respect to the mass. There is a great
deal of evidence to support the presence of this bias at high
redshift. The correlations of Lyman-break galaxies at $z\simeq 3$
(Steidel et al. 1999) are
almost identical to to those of present-day field galaxies, even
though the mass must have been much more uniform at early
times. Furthermore, the correlations increase with UV luminosity
(Giavalisco \& Dickinson 2001) reaching scale lengths of $r_{0} \simeq
7.5 h^{-1}$ Mpc - approximately 1.5 times the present-day value. In
the case of luminous proto-ellipticals we expect an even stronger bias
since we select not just massive galaxies but those that have
collapsed particularly early in order to generate the oldest stellar
populations. This is suggested by studies of the local Universe which
have shown that early-type galaxies are much more clustered than
late-type galaxies (eg. Guzzo et al. 1997, Willmer et al. 1998), and
more recently by the findings of Daddi et al. (2000) who have
investigated the clustering properties of extremely red objects (EROs). They
detect a strong clustering signal of the EROs which is about an order
of magnitude larger than the  clustering of $K-$selected field
galaxies, and also  report a smooth trend of increasing clustering
amplitude with increasing $R-K$ colour, reaching $r_{0} \simeq
11 h^{-1}$ Mpc for $R-K>5$. These results are probably the strongest
evidence to date that the largest fraction of EROs is composed of
ellipticals at $z>1$.

We already have some hints of the possibility of strong clustering 
in the bright sub-mm population from the discovery of a strong excess of 
bright SCUBA sources around high-redshift AGN (Ivison et al. 2000). 
Motivated by this result, and the potential connection with the 
intermediate-redshift EROs discussed above, we have attempted to quantify the 
possible strength of clustering in our two survey fields via calculation
of the the two-point correlation function. The results, based purely on 
sources with $S/N > 3.50$, are shown in Figs 12 and 13.
There is tentative evidence of clustering on scales of $1-2$
arcmin in both of our survey fields, but most particularly, and
somewhat stronger, in the ELAIS N2 region. Referring to Fig 9, we
can in fact see by eye that the most 
significant ($S/N > 3.50$) sources in ELAIS N2 do not conform to a
homogeneous distribution across the field - rather there are two apparent
concentrations of 850~$\rm \mu m$ sources in the top left and bottom right
of the image, approximately at RA 16:37:00, DEC +41:05:00, and RA
16:36:30, DEC +40:56:00 respectively, with an apparent under-density of
sources in the intervening regions. A second peak in $w(\theta)$ at
$\simeq 600$ arcsec (the distance between these over-densities),
and a trough at $\simeq 400$ arcsec are seen to reflect this ``by
eye'' distribution. Very interestingly, we see the same large-scale 
inhomogeneities
in the Chandra X-ray image of the ELAIS N2 region (Almaini et
al. 2002), although the coincidence of X-ray and SCUBA sources is
small ($<10\%$). Almaini et al. provide a full discussion and propose
a possible explanation for this effect.

It is clear from Figs 12 and 13 that, due to small-number statistics,
this first attempt at a direct measure of the clustering of bright sub-mm
sources has proved inconclusive. However, this should not be taken as evidence
that the SCUBA sources are unclustered. In fact, as illustrated in these
figures, it is worth noting that 
these correlation functions are certainly still consistent with
the strong clustering signal detected for EROs by Daddi et al. (2000).
Clearly a much larger survey
($\sim 0.5$ square degrees) will be required to obtain a meaningful
measurement of the strength of clustering in the bright sub-mm population.
Near complete redshift information may also be required to quantify
the extent to which any clustering signal
will be partially erased by projection through a 
wide range in redshift. For example, suppose that 
the bright SCUBA population spans a relatively wide range 
in redshift $2 < z < 5$ whereas, as argued above, the bright EROs are 
confined to the redshift range $1 < z < 2$. In that case, if the strength 
of the spatial clustering in the two populations was the same, $w(\theta)$
as measured from SCUBA images would be expected to be $1.6 - 1.9$ times
smaller (depending on choice of cosmology) 
than that which has been measured for EROs.

\section{Conclusions}

We have used the SCUBA camera on the JCMT to conduct 
the largest extra-galactic sub-mm survey to date, 
in two regions of 
sky (ELAIS N2 and Lockman Hole East)
covering a total area of 260~$\mathrm{arcmin^{2}}$ to a typical rms
noise level of $\sigma_{850} \simeq 2.5$~mJy/beam. We have reduced 
the resulting dataset using both the standard JCMT SURF software and our own
IDL-based pipeline which allows us to produce zero-footprint maps and
noise images. The uncorrelated noise maps from the latter approach
have enabled us to quantify the statistical significance of each peak
in our maps leading to properly quantified errors on the flux density
of our sources. We find 19 sources with $S/N > 4.00$, 38 sources with $S/N > 3.50$ and 72 sources
with $S/N > 3.00$. A series of Monte-Carlo simulations have allowed us
to assess the completeness, positional uncertainties and effects of
confusion in this survey, 
leading to the most accurate estimate of the sub-millimetre
counts over the flux-density range $S_{850} = 5-12$~mJy. 
Our best estimate of the cumulative source count
at $S_{850}>8$~mJy is $320^{+80}_{-100}$, consistent with a range 
of models involving strong evolution of the dust-enshrouded 
starburst population.

Adopting $\Omega_M = 0.3$, $\Omega_{\Lambda} = 0.7$ and 
$H_0 = 67$~${\rm kms^{-1}Mpc^{-1}}$, the sources uncovered by this survey 
have inferred star-formation rates in excess of 
$1000 {\rm M_{\odot}yr^{-1}}$, enshrouded in $10^{8} - 10^{9}~{\rm M_{\odot}}$ of dust. Assuming simply that 
the majority of these sources lie at $z>1.5$, this result implies that
their co-moving number density is $\simeq 10^{-5} {\rm Mpc^{-3}}$, 
coincident with the number density of massive ellipticals with $L >
3-4 L^{\star}$ in the present-day Universe. This co-moving 
number density is also the same as
that inferred for comparably massive
passively-evolving objects in the redshift band $1<z<2$ from recent
surveys of EROs. Thus while the bright submillimetre sources uncovered
by this survey contribute only $\simeq 10\%$ of the submillimetre
background, they can plausibly account for the formation of all
present-day massive ellipticals. A key test of this picture is to determine
whether the bright SCUBA population is at least as strongly clustered as 
are the EROs at intermediate redshifts. As a first attempt at such a test
we have compared the distribution
of our sources with a mock catalogue of fake sources. Our results have proved
inconclusive due to the small number of significant sources
(S/N$>3.50$) detected in our survey areas, but there are some initial
indications of clustering on scales of $1-2$ arcminutes, particularly
in the ELAIS N2 field. 

The process of multi-wavelength follow-up of this survey is already well 
underway. SED-based redshift constraints, along with potential
optical-infrared identifications of
our most significant sources are presented in paper II (Fox et al. 2002),
while the results of IRAM PdB interferometer imaging of 
the most significant Lockman Hole East is presented by Lutz et al. (2001). 
Correlations with deep X-ray and radio images will also be discussed in 
subsequent papers by Almaini et al. (2002) and Ivison et al. (2002).

\section*{ACKNOWLEDGEMENTS}

Susie Scott acknowledges the support of a PPARC Studentship, while 
James Dunlop acknowledges the enhanced research time provided by the 
award of a PPARC Senior Fellowship. We are
very grateful to the many members of the Joint Astronomy Centre for
their continued help and support of this project. We also thank Steve
Eales and Amy Barger for communicating their source count values, and
Omar Almaini for many useful discussions. The JCMT
is operated by the Joint Astronomy Centre on behalf of the UK Particle
Physics and Astronomy Research Council, the Canadian National Research
Council and the Netherlands Organization for Scientific Research. We
also thank our anonymous referee for useful suggestions.

\newpage
\appendix
\onecolumn

\section{THE SOURCE-EXTRACTION ALGORITHM.}

Suppose we consider a normalised beam-map $B(x,y)$ as a source template
and that at position $(i,j)$ in the unconvolved image the signal is
$S(i,j)$ and the noise is $N(i,j)$. If n peaks above a specified flux
threshold are located in the Gaussian-convolved image, we may construct a
model to the unconvolved image such that beam-maps centred on
each peak position are simultaneously scaled to give an overall best
fit to the entire image. Using a minimised $\chi^{2}$ fit as the
maximum likelihood estimator then \be
\chi^{2} = \left( \sum_{i,j} \frac{S(i,j) - \sum_{k=1}^{n} a_{k} B_{k}(x-i,y-j)}{N(i,j)^{2}} \right)^{2}
\ee where $a_{k}$ is the best fit flux to the $k$th peak. Minimising with
respect to each $a_{k}$
\be
\frac{d\chi^{2}}{da_{m}} =
2 \sum_{i,j}\frac{[S(i,j) - \sum_{k=1}^{n} a_{k}
B_{k}(x-i,y-j)] B_{m}(x-i,y-i)}{N(i,j)^{2}} = 0
\ee
and $m= 1 \cdots n$. \\
Rearranging, we obtain the matrix equation
\be
\sum_{k=1}^{n} \left( \sum_{i,j} \frac{B_{k}(x-i,y-j)
B_{m}(x-i,y-j)}{N(i,j)^{2}} \right) a_{k} = \sum_{i,j} \frac{S(i,j)
B_{m}(x-i,y-j)}{N(i,j)^{2}}
\ee
which may be written in the form
\be
\sum_{k=1}^{n} \alpha_{mk} a_{k} = \beta_{m}
\ee
where
\be
\alpha_{mk}= \sum_{i,j} \frac{B_{k}(x-i,y-j)
B_{m}(x-i,y-j)}{N(i,j)^{2}}
\ee
an $n \times n$ matrix, and 
\be
\beta_{m} = \sum_{i,j} \frac{S(i,j)
B_{m}(x-i,y-j)}{N(i,j)^{2}}
\ee
a vector of length n. \\
To recover the best fit values of $a_{k}$ we invert matrix $\alpha$
such that
\be
a_{k} = \sum_{m=1}^{n} [\alpha]_{km}^{-1} \beta_{m}
\ee
The variance associated with the estimate $a_{k}$ is given by
\be 
N^{2}(a_{k}) = \sum_{i,j} N(i,j)^{2} \left(\frac{\partial
a_{k}}{\partial S(i,j)} \right)
\ee
Since $\alpha_{km}$ is independent of $S(i,j)$
\be
\frac{\partial a_{k}}{\partial S(i,j)} = \sum_{m=1}^{n}
\frac{[\alpha]_{km}^{-1} B_{m}(x-i,y-j)}{N(i,j)^{2}} 
\ee
Consequently we find,
\be
N^{2}(a_{k}) = \sum_{k=1}^{n} \sum_{l=1}^{n} [\alpha]_{km}^{-1}
[\alpha]_{kl}^{-1} \left( \sum_{i,j} \frac{B_{m}(x-i,y-j)
B_{l}(x-i,y-j)}{N(i,j)^{2}} \right)
\ee
The final term in brackets is simply the matrix $[\alpha]$ and so this
expression reduces to
\be
N^{2}(a_{k}) = [\alpha]_{kk}^{-1}
\ee
The diagonal elements of $[\alpha]^{-1}$ are the variances of the
fitted parameters $a_{k}$ such that the significance of the peak
detection
\be
\sigma(a_{k}) = \frac{a_{k}}{\sqrt{[\alpha]_{kk}^{-1}}}
\ee

\begin{thebibliography}{80}

\bibitem{1} Almaini, O., et al., 2002, MNRAS in press (astro-ph/0108400).
\bibitem{2} Almaini, O., Pettini, M., Steidel, C.C., 2002, in prep.
\bibitem{3} Archibald, E.N., Dunlop, J.S., Hughes, D.H., Rawlings, S., Eales,
S.A., Ivison, R.J., 2000a, MNRAS, 323, 417.
\bibitem{4} Archibald, E.N., Wagg, J.W., Jenness, T., 2000b, 
{\tt http://www.jach.hawaii.edu/JACdocs/JCMT/SCD/SN/002.2/}
\bibitem{74} Archibald, E.N., Dunlop, J.S., Jimenez, R., Friaca,
A.C.S., Mclure, R.J., Hughes, D.H., 2001, ApJ, submitted (astro-ph/0108122).
\bibitem{7} Baker, A.J., Lutz, D., Genzel, R., Tacconi, L.J.,
Lehnert, M.D., 2001, A\&A, in press (astro-ph/0104345).
\bibitem{5} Barger, A.J., Cowie, L.L., Sanders, D.B., Fulton, E., Taniguchi, Y, Sato, Y., Kawara, K., Okuda, H., 1998, Nat, 394, 248. 
\bibitem{6} Barger, A.J., Cowie, L.L., Sanders, D.B., 1999, ApJ, 518, L5. 
\bibitem{78} Barger, A.J., Cowie, L.L., Smail, I., Ivison, R.J.,
Blain, A.W., Kneib, J.-P., 1999, AJ, 117, 2656.
\bibitem{76} Barger, A.J., Cowie, L.L., Richards, E.A., 2000, AJ, 119, 2092.
\bibitem{8} Baugh, C.M., Cole, S., Frenk, C.S., 1996, MNRAS, 282, L27.
\bibitem{9} Blain, A.W., Kneib, J.-P., Ivison, R.J., Smail, I., 1999, ApJ, 512,
L87. 
\bibitem{72} Borys, C., Chapman, S., Halpern, M., Scott, D., 2002, MNRAS, 330, L63.
\bibitem{10} Coulson, I., 2000, {\tt http://www.jach.hawaii.edu/JACpublic/
JCMT/Facility\_description/Pointing/problems.html\#new} 
\bibitem{11} Cowie, L.L., Hu, E.M., Songaila, A., 1995, Nat, 377, 603.
\bibitem{73} Cowie, L.L., Songaila, A., Barger, A.J., 1999, AJ, 118, 603.
\bibitem{12} Daddi, E., Cimatti, A., Pozzetti, L., Hoekstra, H., Röttgering,
H.J.A., Renzini, A., Zamorani, G., Mannucci, F., 2000, A\&A, 361, 535.
\bibitem{13} Daddi, E., Cimatti, A., Renzini, A., 2002, A\&A, in press (astro-ph/0010093).
\bibitem{14} Dunlop, J.S., 2001a, in: FIRSED2000, eds. I.M. van
Bemmel, B. Wilkes, \& P. Barthel, Elsevier New Astronomy Reviews, in
press (astro-ph/0101297).
\bibitem{15} Dunlop, J.S., 2001b, in: QSO Hosts and their Environments, 
eds. Marquez, I. et al. Kluwer, in press. (astro-ph/0103238).
\bibitem{16} Dunlop, J.S., 2001c, in: Deep Sub-millimetre Surveys, eds.
Lowenthal, J. \& Hughes, D.H.,
World Scientific, in press. (astro-ph/0011077)
\bibitem{17} Dunlop, J.S., Hughes, D.H., Rawlings, S., Eales, S.A., Ward, M.J.,
1994, Nat, 370, 347.
\bibitem{18} Dunlop, J.S., McLure, R.J., Kukula, M.J., Baum, S.A., O'Dea, C.P., Hughes, D.H., 2002, MNRAS, in press.
\bibitem{19} Dunne, L., Eales, S., Edmunds, M., Ivison, R., Alexander,
P., Clements, D.L., 2000, MNRAS, 315, 115.
\bibitem{20} Dunne, L., Eales, S., 2001, 327, 697.
\bibitem{21} Eales, S., Lilly, S., Webb, T., Dunne, L., Gear, W., Clements, D.,
Yun, M., 2000, AJ, 120, 2244.
\bibitem{22} Elbaz, D., Cesarsky, C.J., Fadda, D., Aussel, H., D$\mathrm{\acute{e}}$sert, F.X.,
Franceschini, A., Flores, H., Harwit, M., Puget, J.L., Starck, J.L.,
Clements, D.L., Danese, L., Koo, D.C., Mandolesi, R., 1999, A\&A, 351, L37.
\bibitem{23} Efstathiou, A., et al., 2000, MNRAS, 319, 1169.
\bibitem{24} Fixsen, D.J., Dwek, E., Mather, J.C., Bennel, C.L., Shafer, R.A.,
1998, ApJ, 508, 123.
\bibitem{25} Fox, M.J., et al., 2002, MNRAS, 331, 839.
\bibitem{77} Fruchter, A.S., Hook, R.N., 2002, PASP, 114, 144.
\bibitem{26} Gardner, J.P., Sharples, R.M., Frenk, C.S., Carrasco, B.E., 1997, ApJ, 480, L99.
\bibitem{27} Giavalisco, M., Dickinson, M., ApJ, 550, 177.
\bibitem{28} Glazebrook, K., Ellis, R., Colless, M., Broadhurst, T.,
Allington-Smith, J., Tanvir, N., 1995, MNRAS, 273, 157.
\bibitem {29} Guzzo, L., Strauss, M.A., Fisher, K.B., Giovanelli, R., Haynes, M.P., 1997, ApJ, 489, 37.
\bibitem{30} Hauser, M.G., et al, 1998, AJ, 115, 1418.
\bibitem{31} Holland, W.S., Robson, E.I., Gear, W.K., Cunningham, C.R.,
Lightfoot, J.F., Jenness, T., Ivison, R.J., Stevens, J.A., Ade,
P.A.R., Griffin, M.J., Duncan, W.D., Murphy, J.A., Naylor, D.A., 1999,
MNRAS, 303, 659.
\bibitem{32} Hughes, D.H., Dunlop, J.S., Rawlings, S., 1997, MNRAS,
289, 766.
\bibitem{33} Hughes, D.H., Dunlop, J.S., Rowan-Robinson, M., Serjeant, S., Blain,
A., Mann, R.G., Ivison, R.J., Peacock, J., Efstathiou, A., Gear, W.,
Oliver, S., Lawrence, A., Longair, M., Goldschmidt, P., Jenness, T., 1998, Nat,
394, 241.
\bibitem{34} Hughes, D.H., Gazta\~naga, 2000, in: Proc. of 33rd ESLAB
Symp. ``Star formation from the small to the large scale'' (F. Favata,
A.A. Kaas \& A. Wilson eds., ESA SP-445, 2000).
\bibitem{35} Ivison, R.J., 1995, MNRAS, 275, L33.
\bibitem{36} Ivison, R.J., Dunlop, J.S., Hughes, D.H., Archibald, E.N., 
Stevens, J.A., Holland, W.S., Robson, E.I., Eales, S.A., Rawlings, S., Dey, A.,
Gear, W.K., 1998, ApJ, 494, 211.
\bibitem{37} Ivison, R.J., Dunlop, J.S., Smail, I., Dey, A., Liu, M.C., 
Graham, J.R., 2000, ApJ, 542, 27. 
\bibitem{38} Ivison, R.J., et al., 2002, MNRAS, in press (astro-ph/0206432).
\bibitem{39} Jameson, A., 2000, PHD thesis, University of Cambridge. 
\bibitem{40} Jarvis, M.J., Rawlings, S., Eales, S.A., Blundell, K.M., Willott,
C.J., 2001, in: QSO hosts and their environments, 
eds. Marquez, I. et al., Kluwer, in press.
\bibitem{41} Jenness, T., Lightfoot, J.F., `SURF - SCUBA User Reduction facility
v1.1 Users Manual', 1997.
\bibitem{42} Jenness, T., 2000, {\tt http://www.jach.hawaii.edu/JACdocs/JCMT/
tr/001/84/index.html}
\bibitem{43} Jimenez, R., Friaca, A.C.S., Dunlop, J.S., Terlevich, R.J., Peacock,
J.A., Nolan, L.A., 1999, MNRAS, 305, L16.
\bibitem{44} Jimenez, R., Padoan, P., Dunlop, J.S., Bowen, D.V., Juvela, M., 
Matteucci, F., 2000, ApJ, 532, 152.
\bibitem{45} Kauffman, G., Charlot, S., 1998, MNRAS, 294, 705.
\bibitem{46} Kawara, K., Sato, Y., Matsuhara, H., Taniguchi, Y., Okuda, H., Sofue,
Y., Matsumoto, T., Wakamatsu, K., Karoji, H., Okamura, S., Chambers,
K.C., Cowie, L.L., Joseph, R.D., Sanders, D.B., 1998, A\&A, 336, L9.
\bibitem{79} Kochanek, C.S., et al. 2001, ApJ, 560, 566.
\bibitem{47} Kormendy, J., Gebhardt, K., 2001, 
in: 20th Texas Symposium on Relativistic Astrophysics,
eds. Martel, H., Wheeler, J.C., AIP, in press (astro-ph/0105230).

\bibitem{48} Lilly, S.J., Le F\`evre, O., Hammer, F., Crampton, D.,
1996, ApJ, 460, L1.
\bibitem{49} Lutz, D., et al., 2001, A\& A, 378, 70.
\bibitem{50} Madau, P., Ferguson, H.C., Dickinson, M.E., Giavalisco, M., Steidel,
C.C., Fruchter, A., 1996, MNRAS, 283, 1388.
\bibitem{51} Magorrian, J., et al., 1998, AJ, 115, 2285.
\bibitem{52} Oliver, S., et al., 2000, MNRAS, 316, 749.
\bibitem{53} Puget, J.L., Abergel, A., Bernard, J.P., Boulanger, F., Burton, W.B.,
Desert, F.X., Hatmann, D., 1996, A\&A, 308, L5.
\bibitem{54} Renzini, A., Cimatti, A., 2000, in The Hy-Redshift Universe: Galaxy Formation and Evolution at High Redshift, Conf. proc., Berkeley 1999, in press (astro-ph/9910162).
\bibitem{55} Rowan-Robinson, M., et al., 1997, MNRAS, 289, 490.
\bibitem{56} Rowan-Robinson, M., 2000, MNRAS, 316, 885. 
\bibitem{57} Saunders, W., Rowan-Robinson, M., Lawrence, A., Efstathiou, G.,
Kaiser, N., Ellis, R.S., Frenk, C.S., 1990, MNRAS, 242, 318.
\bibitem{58} Sawicki, M., 2000, A\&AS, 197, 6504.
\bibitem{59} Scoville, N.Z., Young, J.S., 1983, ApJ, 265, 148.
\bibitem{60} Serjeant, S., et al., 2000, MNRAS, 316, 768.
\bibitem{61} Serjeant, S., et al., 2002, MNRAS, in press (astro-ph/0201502).
\bibitem{62} Smail, I., Ivison, R.J., Blain, A.W., 1997, ApJ, 490, L5.
\bibitem{63} Smail, I., Ivison, R.J., Blain, A.W., Kneib, J.-P., 2002, MNRAS, 331, 495.
\bibitem{64} Steidel, C.C., Adelberger, K.L., Giavalisco, M., Dickinson, M.,
Pettini, M., 1999, ApJ, 519, 1.
\bibitem{65} Szokoly, G.P., Subbarao, M.U., Connolly, A.J., Mobasher, B., 1998,
ApJ, 492, 452.
\bibitem{66} Thronson, H., Telesco, C., 1986, ApJ, 309, L79.
\bibitem{67} van der Werf, P.P., Kraiberg Knudsen, K., Labb\'e, I.,
Franx, M., 2001,in: Deep Sub-millimetre Surveys, eds.
Lowenthal, J. \& Hughes, D.H.,
World Scientific, in press. (astro-ph/0010459). 
\bibitem{68} Wang, B., 1991, ApJ, 374, 456.
\bibitem{69} Warren, S.J., Hewett, P.C., Osmer, P.S., 1995, ApJ, 438, 506.
\bibitem{70} Willmer, C.N.A., da Costa, L.N., Pellegrini, P.S., 1998, AJ, 115, 869.
\bibitem{71} Willott, C.J., Rawlings, S., Blundell, K.M., 
2001, MNRAS, 324, 1.
\end{thebibliography}
\end{document}